\documentclass[twocolumn,twocolappendix,tighten,appendixfloats]{aastex6}
\usepackage{natbib}
\usepackage{amssymb}
\usepackage{multirow}
\usepackage{graphicx}
\usepackage{tikz}

\newcommand{\mk}{\rm Mrk~71}
\newcommand{\hi}{H {\small I}}
\newcommand{\hii}{H {\small II}}
\newcommand{\heii}{He {\small II}}
\newcommand{\oiii}{[O {\small III}]}
\newcommand{\oii}{[O {\small II}]}
\newcommand{\lyc}{LyC}
\newcommand{\hb}{H$\beta$}
\newcommand{\ngc}{NGC 2366}
\newcommand{\ha}{H$\alpha$}

\begin{document}
\submitted{Submitted: 3 March 2017}
\accepted{24 July 2017}

\title{Mrk 71 / NGC 2366:  The Nearest Green Pea analog}


\author{Genoveva Micheva\altaffilmark{1}, M. S. Oey\altaffilmark{1}, Anne E. Jaskot\altaffilmark{2}, and Bethan L. James\altaffilmark{3}}



\altaffiltext{1}{University of Michigan, 311 West Hall, 1085 S. University Ave, Ann Arbor, MI 48109-1107, USA}
\altaffiltext{2}{Department of Astronomy, Smith College, Northampton, MA 01063, USA}
\altaffiltext{3}{STScI, 3700 San Martin Drive, Baltimore, MD 21218, USA}

\begin{abstract}
We present the remarkable discovery that the dwarf
irregular galaxy NGC~2366 is an excellent analog of the Green Pea (GP)
galaxies, which are characterized by extremely high ionization parameters.
The similarities are driven predominantly by the giant \hii\ region
Markarian 71 (\mk).  We compare the system with GPs 
in terms of morphology, excitation properties, specific star-formation
rate, kinematics, absorption of low-ionization species, reddening, and
chemical abundance, and find consistencies throughout.  Since extreme
GPs are associated with both candidate and confirmed Lyman continuum
(\lyc) emitters, \mk/\ngc\ is thus also a good candidate for
\lyc\ escape.  The spatially resolved data for this object
show a superbubble blowout generated by mechanical feedback from one
of its two super star clusters (SSCs), Knot B, while the extreme
ionization properties are driven by the $\lesssim1$ Myr-old,
enshrouded SSC Knot A, which has $\sim10$ times higher ionizing
luminosity.  Very massive stars ($> 100\ \rm M_\odot$) may be present
in this remarkable object.  Ionization-parameter mapping indicates the
blowout region is optically thin in the \lyc, and the general
properties also suggest \lyc\ escape in the line of sight. \mk/\ngc\ does differ from GPs in that it is 1 -- 2 orders of magnitude
less luminous.  The presence of this faint GP analog and candidate
\lyc\ emitter (LCE) so close to us suggests that LCEs may be
numerous and commonplace, and therefore could significantly contribute
to the cosmic ionizing budget. \mk/\ngc\ offers an unprecedentedly
detailed look at the viscera of a candidate LCE, and
could clarify the mechanisms of LyC escape. 
\end{abstract}

\keywords{}

\section{Introduction} \protect\label{sec:intro}
Lyman continuum (\lyc) radiation from star-forming (SF) galaxies escapes into the intergalactic medium (IGM) and contributes to the ionizing budget of the Universe. A key point of interest is to identify the mechanisms of escape and how the properties of the IGM and the emitting galaxy govern the \lyc\ escape fraction ($f_{\rm esc}$).  Ideally, this requires the detection of \lyc\ emitters (LCEs), located close enough to be spatially resolved and studied in detail, as well as statistically large LCE samples from galaxy populations at different redshifts.  

At low redshifts the direct detection of LCEs has been notoriously difficult, resulting mostly in upper limits~\citep[e.g.][]{1995ApJ...454L..19L,2001ApJ...546..665S, 2007ApJ...668...62S,2015ApJ...804...17S,2010ApJ...725.1011V}, with a handful of confirmed \lyc\ detections:  Haro 11~\citep[$f_{\rm esc}\sim3.3\%$;][]{2011A&A...532A.107L}, Tol 1247-232~\citep[$f_{\rm esc}\sim4.5\%$;][]{2013A&A...553A.106L,2016ApJ...823...64L}, Mrk 54~\citep[$f_{\rm esc}\sim2.5\%$;][]{2016ApJ...823...64L}, and J0921+4509~\citep[$f_{\rm esc}\sim1\%$;][]{2014Sci...346..216B}.
At high redshifts statistically significant samples are easier to obtain; however, the problem of foreground contamination and the severe attenuation of \lyc\ by the IGM has resulted in many non-detections~\citep[e.g.][]{2015ApJ...804...17S,2015A&A...576A.116V,2016A&A...587A.133G,2017MNRAS.465..316M}, with few confirmed LCEs~\citep[][]{2016A&A...585A..51D,2016ApJ...826L..24S}. 

Recently, a new subclass of local compact emission-line galaxies, the Green Peas~\citep[GPs;][]{2009MNRAS.399.1191C}, has been proposed, not only as an excellent analog class of high redshift strongly star-forming galaxies like Lyman alpha emitters (LAEs), but also as strong candidates for \lyc~emission~\citep{2013ApJ...766...91J}.  That the extreme GPs are viable LCE candidates has been dramatically confirmed by~\citet{2016Natur.529..178I,2016MNRAS.461.3683I}, who spectroscopically observed five GPs at redshifts $z\sim0.3$ with the Hubble Space Telescope (HST).  In all five of their target objects, they directly detect strong \lyc~emission in the range $f_{\rm esc}=6\mbox{--}13\%$. These are the highest escape fractions measured to date among low-redshift SF galaxies. Note that with the detections by Izotov et al., the number of confirmed \lyc~leakers at low redshifts has now doubled. The GP class therefore offers a plethora of strong LCE candidates.

\begin{figure*}
\plotone{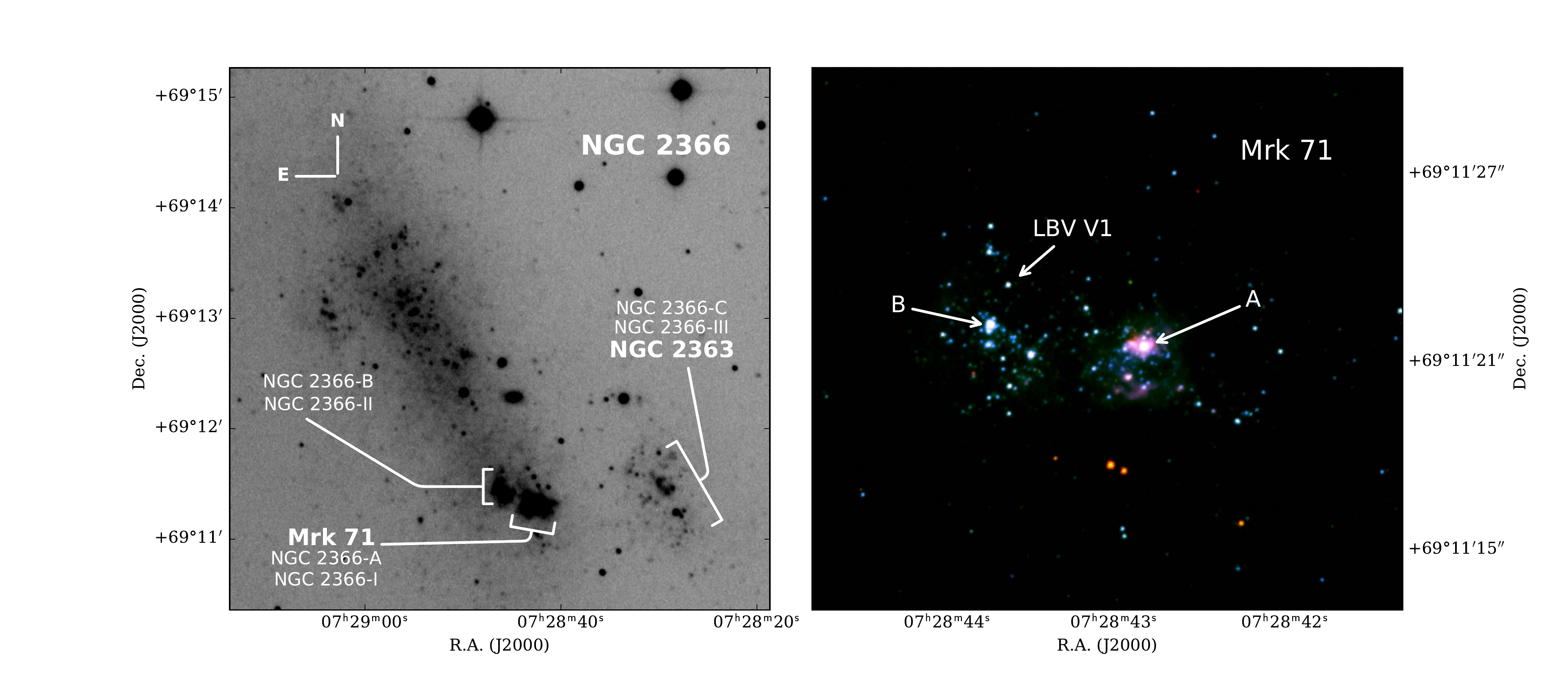}
\caption{Left: $R$-band image~\citep{2009ApJ...703..517D} of the NGC
  2366 system, including NGC 2366 and neighboring galaxy NGC~2363.
  The giant \hii\ region \mk\ is identified at the south end of
  NGC~2366.  Mrk~71 is often erroneously referred to as ``NGC 2363'' or ``NGC 2363-A''.
  Some designations from the literature are shown:  NGC
  2366-I, -II, -III are from~\citet{2000AJ....119..688D}, and NGC
  2366-A, -B, -C are from~\citet{1991ApJ...367..141R}.  Right:  Color
  composite from archive {\it HST}/WFC3 broad band imaging of
  \mk\ \citep{2016ApJ...816...40J} with
  (red, green, blue) = (F814W, F547M, F336W), showing the positions of
  Knots A and B. \protect\label{fig:nomenclature}}  
\end{figure*}

GPs are a rare class of low-metallicity, vigorously star-forming
galaxies~\citep{2009MNRAS.399.1191C,Izotov2011}. In the ``BPT''
diagram of~\citet{1981PASP...93....5B} with ${\rm log}([{\rm N
    II}]/{\rm H}\alpha)$ vs. ${\rm log}([{\rm OIII}]/{\rm H}\beta)$
they occupy the most extreme end of low-mass, low-metallicity local SF
galaxies in the Sloan Digital Sky survey (SDSS). They are defined by
[OIII]$\lambda 5007$ emission-line equivalent widths exceeding
hundreds of \AA~\citep{2009MNRAS.399.1191C}. At their typical redshift
of $z\sim0.2$ this emission falls into the Sloan $r$ band and
dominates the luminosity of the 3-color $gri$ image. They have
sub-solar
metallicities~\citep{2010ApJ...715L.128A,2012ApJ...749..185A,2012PASP..124...21H},
compact morphology, and some of the highest specific star formation
rates (sSFR) in the local Universe.~\citet{2015ApJ...809...19H} show
that GPs are similar to high redshift galaxies in terms of having
double-peaked Ly$\alpha$ profiles, with velocity separations
indicative of low HI column densities, and Ly$\alpha$ luminosities and
equivalent widths in the range of most high redshift LAEs. GPs have
extremely high ionization parameters, as indicated by their
[OIII]$\lambda5007$/[O II]$\lambda3727$ ratios, which are some of the
highest among local galaxies. In terms of ionization properties GPs
are therefore comparable to high redshift LAEs~\citep{2014MNRAS.442..900N}.  
GPs are therefore a vital population for understanding which galaxies and
which physical conditions are responsible for the release of LyC
radiation into the intergalactic medium, and ultimately, for cosmic
reionization. 

We have made a serendipitous discovery of the nearest LCE candidate and GP analog, conveniently located at only $3.4$ Mpc~\citep{1995AJ....110.1640T}. This object is the dwarf irregular galaxy \ngc, whose star formation is strongly dominated by the giant HII region Markarian 71 (\mk). This system is not only consistent with various properties of the GP class, but also is often more extreme than known LCE GPs, which suggests that it is a strong LCE candidate. \ngc\ is close enough to study in great detail, at a level unprecedented for other GPs or even local LCEs, since it is $25$ times closer than the nearest confirmed LCE (Haro 11 at $84.3$ Mpc). Given the spatial detail with which we can examine this object, \mk\ may clarify mechanisms for \lyc\ escape.  In this introductory paper, we establish the similarities between the GPs and \ngc, with a focus on the dominant HII region \mk, as well as examine the likelihood of \lyc\ escape.  We will examine the physical processes in \mk\ in a subsequent work. 

\subsection{Nomenclature}\protect\label{sec:nomenclature}
NGC 2366 is a Magellanic barred irregular galaxy of
class IB(s)m~\citep{1991rc3..book.....D}. There is unfortunate confusion with multiple, and
erroneous, nomenclature relating to \mk\ and its neighboring complexes, so 
in Figure~\ref{fig:nomenclature} we present NGC 2366
with the main substructures indicated with their correct
designations. In the left panel of the figure, the large, luminous HII
region complex located to the south is \mk.  The structure to the west of
\mk~is another dwarf galaxy, NGC 2363, interacting with NGC~2366.  The right panel of
Figure~\ref{fig:nomenclature} shows a zoom on \mk, with its two
prominent super star clusters (SSCs), named A and B by~\citet{1994ApJ...437..239G},
indicated in the figure.  We will refer to these identifications for
all of our work.

Unfortunately, \mk\ has often been mis-identified as NGC 2363 \citep[e.g.,][]{1980PASP...92..134K,1994ApJ...437..239G,1997ApJS..108....1I,2000AJ....119..688D,2001ApJ...556..773H,2011AJ....141...37L}, whereas in fact, the latter is the neighboring galaxy to the west of \mk~in Figure~\ref{fig:nomenclature}a \citep{Corwin2006}.  This mislabeling has propagated into the naming of the two SSCs in \mk, i.e., clusters A and B are sometimes called ``NGC 2363-A'' and ``NGC 2363-B'' \citep[e.g.,][]{1994ApJ...437..239G,2000AJ....119..688D,2011AJ....141...37L}.

To aid cross-referencing, we also indicate alternative
designations in Figure~\ref{fig:nomenclature}.  The three most
prominent substructures associated with NGC 2366 have been
labeled NGC 2366-I ($=$\mk), NGC 2366-II (bright HII region
to the north-east of \mk), and NGC 2366-III ($=$ NGC 2363)
by~\citet{2000AJ....119..688D}.  These same structures were
instead referred to as NGC 2366-A, NGC 2366-B, and NGC 2366-C
by~\citet{1991ApJ...367..141R}.

Figure~\ref{fig:nomenclature}
is representative of our attempt to revert to the original
nomenclature in order to clear up the confusion around the different
naming conventions used in the literature. We will refer to the parent
galaxy only as NGC 2366, to its brightest HII region only
as \mk, and to the two SSCs in \mk\ as Knot A and Knot B
throughout this paper.  We refer to the entire combined ensemble of
NGC 2366 including Mrk 71 and NGC 2366-II, plus NGC 2363, as the ``NGC
2366 system''.



\section{\ngc~as a green pea analog}\protect\label{sec:bigGP}
The HII region \mk~has been extensively studied in the literature both on its own
merit and as part of the larger host
galaxy
\ngc~\citep[e.g.][]{Masegosa1991,1992ApJ...386..498R,1994ApJ...437..239G,1997ApJS..108....1I,2000AJ....119..688D,2000A&A...361...33N,2011ApJ...734...82I,2014MNRAS.441.1841T,2016ApJ...816...40J}.
We will demonstrate that this system shares the key
properties of GPs and Lyman-break analogs (LBAs), making it a unique
local analog of GPs, as well as a strong LCE candidate. To facilitate the comparison
with these galaxy classes, we define these comparison samples below. 

\textbf{The ``average'' GP sample:} GPs were first introduced
by~\citet{2009MNRAS.399.1191C}, who selected the galaxies by eye from
the Galaxy Zoo forum to be ``compact'' and ``green'' in the SDSS $g$,
$r$, $i$ composite images. The properties of this sample that emerged
through their analysis of SDSS spectra revealed an average redshift of
$z\sim0.2$, extremely high equivalent width of [OIII]
$\lambda\lambda4959,5007$, sub-solar metallicities, low masses, and
high star formation rates. This sample has been subsequently well
studied and re-analysed by, for example,~\citet{Izotov2011}, who
provide updated stellar masses;~\citet{2012PASP..124...21H}, who give
detailed abundances; and~\citet{2015ApJ...809...19H} who examine the
Ly$\alpha$ properties.  We refer to properties of the Cardamone
samples as those of the ``average'' GP sample throughout this
paper. The wide range of ionization properties of the average GPs
implies that some but not all GPs in this sample are expected to be
LCE candidates.  

\textbf{The ``extreme'' GP sample:} \citet{2013ApJ...766...91J}
suggest that GPs with extreme ionization properties are good
candidates for \lyc~emission. They assemble a subsample of extreme GPs
characterized by high ratios of $[$O III$]/[$O II$]\gtrsim 7$, and
suggest that such high ratios result from either an unusually high
ionization parameter, or alternatively, a less extreme ionization
parameter in combination with low optical
depth. While~\citet{Stasinska2015AA} show that high line ratios
alone do not necessarily imply massive escape of \lyc, the discovery
of five \lyc~leaking GPs through direct detection
by~\citet{2016Natur.529..178I,2016MNRAS.461.3683I} does indeed suggest
that GPs with extreme ionization properties make for excellent LCE
candidates. Throughout this paper we refer to the sample of
``extreme'' GPs as that defined by the properties of the GPs in
\citet{2013ApJ...766...91J}, combined with the five confirmed
\lyc~emitting GPs of~\citet{2016Natur.529..178I,2016MNRAS.461.3683I}. 

\begin{figure}
\plotone{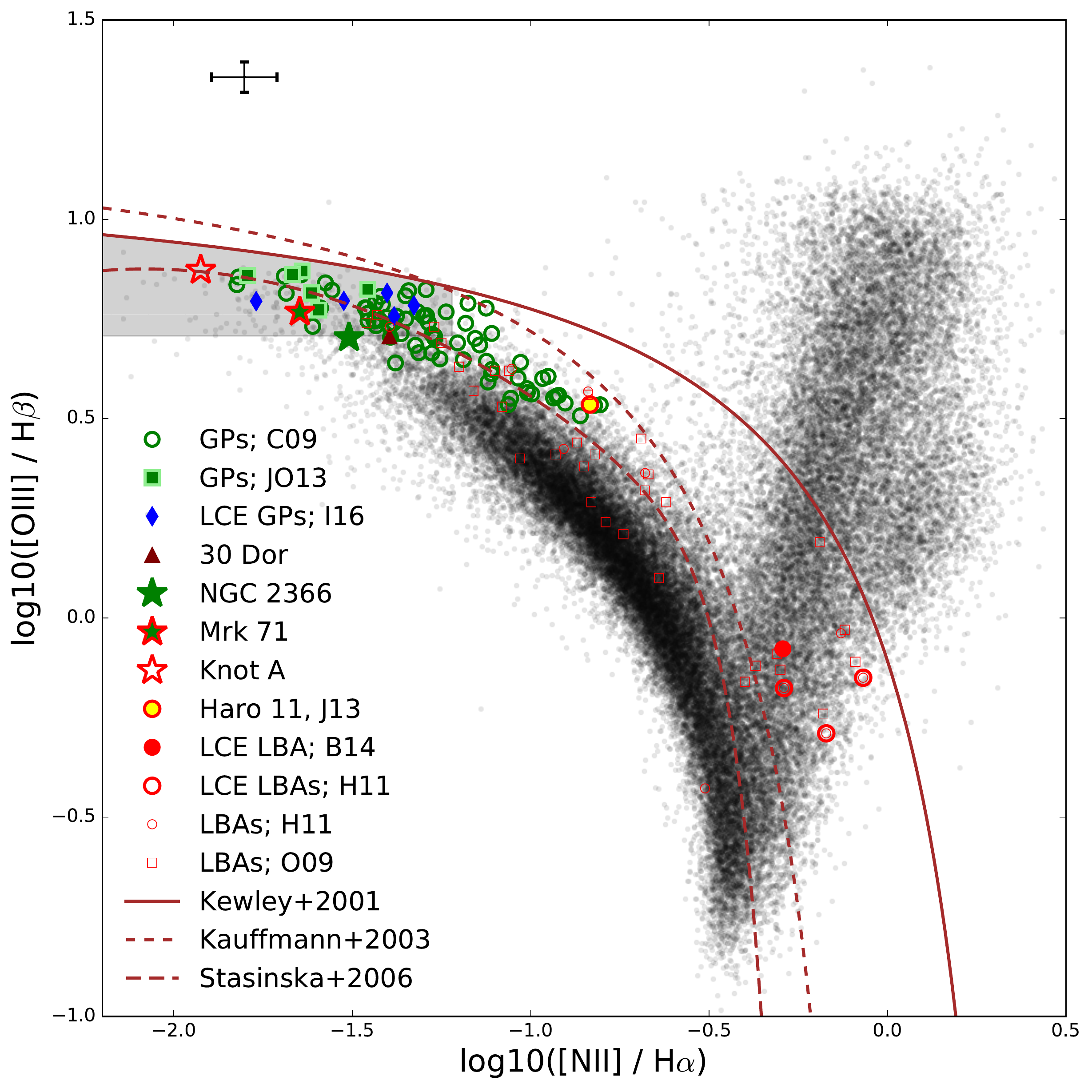}
\caption{BPT diagram showing the ``average GPs'' from~\citet[][C09]{2009MNRAS.399.1191C} with open green circles, the ``extreme GPs'' which are also LCE candidates from~\citet[][JO13]{2013ApJ...766...91J} with filled green squares, the confirmed LCE GPs from~\citet[][I16]{2016Natur.529..178I,2016MNRAS.461.3683I} with filled blue diamonds, LBAs from~\citet[][O09]{2009ApJ...706..203O} with open red squares and those from \citet[][H11]{2011ApJ...730....5H} with thin open red circles. The three unconfirmed LCE LBAs are marked with thick open red circles and the confirmed LCE LBA from~\citet{2014Sci...346..216B} with a filled red circle. 30 Dor~\citep[][]{2003ApJ...584..735P} and the LCE Haro 11~\citep[data from][J13]{2013MNRAS.430.2097J} are shown for comparison. The extreme GPs occupy the far end of the low-mass low-metallicity star-forming branch, and are used to define the filled gray region of extreme GP properties. The size of this region includes the uncertainties of the measurements.\protect\label{fig:BPT}}  
\end{figure}

\textbf{The LBA sample:} Another class of local analogs to star-forming high-redshift galaxies are the LBAs~\citep[][]{2005ApJ...619L..35H,2009ApJ...706..203O,2010ApJ...710..979O}, selected to be the most UV luminous ($L_{\sc \rm FUV}>10^{10.3}L_{\sun}$), and most compact ($I_{\sc \rm FUV}>10^{9} L_{\sun} \textrm{kpc}^{-2}$) star-forming galaxies at redshift $z<0.3$~\citep[e.g.][]{2005ApJ...619L..35H,2007ApJS..173..441H,2008ApJ...677...37O}. LBAs are of interest because some have been confirmed as LCEs~\citep{2014Sci...346..216B}, while others are good LCE
candidates~\citep{2011ApJ...730....5H}.  While they are on average more massive (stellar mass $M_\star=1$--$50\times10^9M_\sun$) and metal rich~\citep[$Z=0.13$--$2.5Z_\sun$, ][]{2001ApJ...558...56H} than GPs, at the metal-poor low-mass end of the LBA distribution they overlap with GPs, and several of the GPs in~\citet{2009MNRAS.399.1191C} are among the LBAs in~\citet{2009ApJ...706..203O}.~\citet{2011ApJ...730....5H} infer a relative $f_{\rm esc}=4$--$12\%$ (assuming a dust-free case) from the residual core intensity of strong UV absorption lines for three of their eight LBAs. This indirect technique of detecting LCE candidates was validated by~\citet{2014Sci...346..216B}, who spectroscopically observe an absolute $f_{\rm esc}=1\%$ from a $z\sim0.2$ LBA, selected from the three candidates in~\citet{2011ApJ...730....5H}. Note that the LCE LBAs are found in the SF-AGN composite region in the BPT diagram (Figure~\ref{fig:BPT}); however, no convincing AGN signatures have been detected in these galaxies~\citep{2009ApJ...706..203O}.  However, reminiscent of the compact nature of GPs, many LBAs are characterized by the presence of a dominant compact object, strong outflow velocities, and high star formation rates.

There is one important distinction between the host galaxy \ngc\ and
these comparison samples.  \ngc\ has stellar mass
$2.6\pm0.3\times10^{8}\ {\rm M_\sun}$~\citep{2014AA...566A..71L}, and
${\rm log}(L_{\sc \rm
  FUV}[L_{\sun}])=8.34$~\citep{2015ApJ...808..109M}, with \mk~likely
dominating this emission. This stellar mass is $\sim4$ times lower
than the typical values in the average GP
sample~\citep[${M_\star}\sim11\times10^8\ {\rm
    M_\sun}$;][]{Izotov2011}, while the luminosity is two orders of
magnitude lower than the average for GPs (${\rm log}(L_{\sc \rm
  FUV}[L_{\sun}])\sim 10.5$;~\citealt{2009MNRAS.399.1191C}). The LBA
sample is brighter and more massive than the average GPs, so the
differences there are even greater. As a GP analog and LCE candidate,
\ngc~therefore probes the previously unexplored region of extremely
faint and low-mass GPs and LCEs. Aside from its much smaller scale, we
now show that the starburst properties of NGC 2366, driven by \mk, are
fully consistent with GPs.   

\mk\ is strongly dominated by two SSCs, Knot A and Knot B.
As seen in Figure~\ref{fig:nomenclature}b, the latter is fully exposed,
and, while the stars are not fully resolved, its stellar population has been
spectroscopically evaluated with both ground-based observations
\citep[][]{1994ApJ...437..239G,2016ApJ...826..194S} and
with {\it HST}/FOS~\citep[][see \S\ref{sec:clusterB}]{2000AJ....119..688D}.
In contrast, the spectrum of Knot A shows no stellar photospheric features at all.  The spectrum is dominated by strong
nebular continuum, including an inverse Balmer break
\citep{1994ApJ...437..239G,2000AJ....119..688D},
suggesting that the SSC in Knot A is still embedded in its natal
cloud. Further supporting evidence comes
from~\citet{2016ApJ...826..194S}, who find that the radio inferred
ionizing flux is considerably larger than that inferred from the
optical, which suggests that the cluster has still not fully
emerged. Nevertheless, Knot A appears to be responsible for most of the
total ionizing luminosity in \mk.  This was shown 
by~\citet{2000AJ....119..688D}, who estimate the number of OB
stars in Knot B from spectral synthesis models of {\it HST}/FOS UV
spectra, obtaining the ionizing photon emission rate per second,
$\log N_{\rm \lyc}=50.78$. This implies an ${\rm H}\alpha$
luminosity for Knot B of $L({\rm H\alpha})=9.1\times10^{38}$ erg
s${}^{-1}$. From H$\alpha$ imaging with {\it HST}/WFC3, \citet{2016ApJ...816...40J} obtain a total luminosity for
\mk\ of $L({\rm H\alpha})=8.4\times10^{39}$ erg s${}^{-1}$. Assuming Case B recombination (but see \S
\ref{sec:mklyc}) or a uniform, single \lyc\ escape fraction between Knots A and
B, Knot B therefore contributes only $\sim11\%$ to the total ionizing budget of \mk,
leaving Knot A as the strongly dominant source for the remaining \lyc\
radiation. As we will show, Knot A also turns out to drive
most of the other features linking \mk, and its parent galaxy \ngc, to GPs and LCE candidates. 

In what follows, we examine a variety of parameters to compare \mk/\ngc\ to
GPs, demonstrating a remarkable correspondence. We summarize these properties in Table~\ref{tab:comparison}.


\subsection{Morphology}\protect\label{sec:morph}
Extreme GPs are compact objects.  At their distances, their bright, dominant starbursts are barely resolved with HST, but in some objects extended structures are detected. The dominant compact regions have typical effective radii of $r_{1/2}=0.3$--$0.7$ kpc (Jaskot et al. in prep). While it appears more extended than  GPs, the \mk\ complex would dominate the luminosity of NGC 2366 if viewed at GP distances, and at wavelengths in which GPs are selected. The apparent diameter of \mk\ along the major axis is $\sim20''\simeq330$ pc. In the redshift range of GPs, $z\simeq0.03$ to $0.3$, \mk~would therefore have an apparent size of $0.6''$ to $0.07''$, respectively, and mimic the appearance of GPs at SDSS resolution. At the same redshifts, the host galaxy NGC 2366 will have $r_{1/2}=4.5''$ and $0.6''$, respectively, with \mk~dominating in brightness in the UV.  From archival GALEX data we measure an NUV surface brightness for \mk, NGC 2366-II, and the nearby dwarf NGC 2363 (Figure~\ref{fig:nomenclature}) 
combined, totaling $5.1\times 10^{-17}\rm erg\ s^{-1}\ cm^{-2}\ \AA^{-1}\ arcsec^{-2}$. This is comparable to the average value for the extended structures that we find around 17 extreme GPs from {\sl HST}/COS NUV acquisition images, $6.6\times10^{-17}\ \rm erg\ s^{-1}\ cm^{-2}\ \AA^{-1}\ arcsec^{-2}$. Over the rest of \ngc~the NUV surface brightness is on average an order of magnitude lower, $5.0\times10^{-18}$, which indicates that, moved to GP redshifts, only the extended region of \mk\ plus NGC 2366-II would likely be visible in the NUV, with the rest of \ngc~below the detection level. This compact appearance is consistent with GPs. Interestingly, even with the remarkable spatial resolution visible in \mk, the dominant object, Knot~A, remains strongly characterized by its compactness.  It is still barely resolved with the HST, and has an estimated upper limit in size of FWHM$<2.3$ pc~\citep{2005ApJ...627..739T} for the compact nebula.

Many LBAs are similarly characterized by a dominant central object~\citep[DCO,][]{2010ApJ...710..979O} with typical $r_{1/2}\sim150$ pc, while the extended host galaxies have a range of optical effective radii $0.48$--$4.6$ kpc~\citep{2009ApJ...706..203O}. NGC 2366 has a $B$-band effective radius consistent with LBAs, $r_{1/2}=2.7\pm0.1$ kpc~\citep{2001ApJ...556..773H}.

\subsection{Excitation}\protect\label{sec:excitation}
One of the primary characteristics of the GPs is their high nebular excitation.  While GPs are identified based on their high \oiii$\lambda$5007 equivalent widths \citep[][see below]{2009MNRAS.399.1191C}, extreme $[\rm OIII]\lambda\lambda5007,4959/[\rm O II]\lambda3727$ line ratios, indicative of correspondingly high ionization parameters, are used as the main characteristic for selecting the extreme GPs in~\citet{2013ApJ...766...91J}. The latter have ratios of $9\mbox{--}14$. The confirmed LCE GPs from~\citet{2016Natur.529..178I,2016MNRAS.461.3683I} have a ratio range of $6.4\mbox{--}8.9$, so with their addition to the extreme sample, our combined comparison regime for this ratio is $6 - 14$, shown in Table~\ref{tab:comparison}. Since this is the defining characteristic of the extreme GP sample, it is of great interest to evaluate this quantity for our local GP analog.  From high resolution HST data for \mk\ of [OIII]$\lambda5007$ and [OII]$\lambda3727$ obtained by \citet{2016ApJ...816...40J}, we estimate [OIII]$\lambda\lambda 5007,4959$/[OII]$\lambda3727=11.7$, taking $\lambda5007/\lambda4959=3.0$. This ratio is well within the extreme GP range. 

To make a first-order estimate of the integrated \oiii/\oii\ for the entire
\ngc\ system, we combine the contributions from the rest of the
components as follows, weighting by the \ha\ luminosity.
The observed [OIII]/[OII]$\sim7.2$ for \mk\ and 2366-II combined, as measured by
\citep{2006ApJS..164...81M} from drift-scanned, ground-based
spectra.  We again corrected for the contribution of $\lambda4959$,
and we additionally corrected for internal reddening using
$C({\rm H}\beta)=0.13$ (\S\ \ref{sec:dust}) because 
the published value  only accounted for Galactic extinction. 
We similarly treated the observed ratio for NGC~2363, which was
also observed by \citet{2006ApJS..164...81M}, obtaining [OIII]$\lambda\lambda5007,4959/{\rm [OII]}=1.08$. 
For the rest of NGC~2366, we adopt mean values of HII regions in the
Small Magellanic Cloud (SMC), since the SMC has a similarly low
metal abundance as \ngc.  For the $10$ \hii\ regions reported by
\citet{1977ApJ...216..706D}, the mean of [OIII]$\lambda\lambda5007,4959$/[OII]$=5.56$. The relative contributions of \mk\ and NGC 2366-II, NGC 2363, and the extended body of \ngc\ to the total \ha\ luminosity are $0.76$, $0.10$, and $0.14$, respectively, as measured from the \ha\ image published by \citet{2009ApJ...703..517D}. Thus, weighting the contributions from all nebular components accordingly, the total mean ratio for the integrated \ngc\ system is [OIII]$\lambda\lambda5007,4959/{\rm [OII]}=6.34$, which is consistent with extreme GPs.  

It is apparent that because \mk\ strongly dominates the \ha\ luminosity of NGC 2366, it similarly dominates the rest of the integrated nebular properties of the system. Therefore, both the integrated \ngc\ system and \mk\ alone are consistent, not only with the average GPs, but also with the extreme GP sample. We note that, on its own, the dominant source of ionization in \mk, Knot A, strongly outperforms the most extreme GPs in our comparison sample, with a ratio of [OIII]/[OII]$=23.0\pm0.6$~\citep{1994ApJ...437..239G}.  Thus, the manifestation of a high excitation ratio depends on both extreme excitation of the dominant starburst and dilution by additional star-forming regions in the unresolved system.  Our values for all the components are given in Table~\ref{tab:comparison}, together with those for the comparison samples.

Further evidence of high ionization is the presence of doubly ionized helium, He II $\lambda4686$, detected both in average GPs \citep[][]{2012PASP..124...21H} and extreme GPs~\citep[][]{2013ApJ...766...91J}, with detection rates of $\sim7\%$ and $\sim50\%$, respectively, and comparable line strengths in the range He II $\lambda4686/{\rm H}\beta \sim 1\mbox{--}2\%$. The extreme GP sample contains the confirmed LCE GPs from~\citet{2016Natur.529..178I,2016MNRAS.461.3683I}; however, these authors do not report He II $\lambda4686$ measurements.
Considering only the six extreme GPs
of~\citep[][]{2013ApJ...766...91J}, for which \heii\ $\lambda4686$ measurements
exist, five out of six of these show positive detections. Thus, the detection
of this line in extreme GPs appears to be more common than among
average GPs.  We note that the line emission is narrow, FWHM$\sim3\mbox{--}5$
\AA~\citep{2013ApJ...766...91J}, and therefore likely nebular in nature, as
expected for conditions with extreme ionization parameter.

For \mk, \citet{2016ApJ...816...40J} obtain He~II $\lambda4686/{\rm H}\beta=0.026$ from {\it HST}/WFC3 narrowband data. 
This emission is largely due to ionization from Knot B, where the presence of Wolf-Rayet (WR) stars has been confirmed~\citep[][\S\ \ref{sec:clusterB}]{2000AJ....119..688D}. Knot A in \mk\ generates He II $\lambda4686/{\rm H}\beta=0.009\pm0.001$~\citep[][]{1997ApJS..108....1I,2016ApJ...816...40J}. For the integrated \ngc\ system, we assume the only signal comes from \mk, with the rest of the system contributing zero to the diluted ratio. Applying the same \ha-based weights as before, we obtain He II $\lambda4686/{\rm H}\beta \sim 0.019$. In terms of the presence and strength of this high-excitation line, \ngc\ is therefore consistent with the extreme GP sample. The nature of the He II $\lambda4686$ emission and the WR population is further discussed in \S~\ref{sec:burstpop}.   

The average GP sample occupies the low-mass, low-metallicity region of the BPT diagram, with line ratios log$([\textrm{OIII}]\lambda5007/\textrm{H}\beta)\gtrsim0.5$ and log$([\textrm{NII}]\lambda6584/\textrm{H}\alpha)\lesssim-0.8$; while the extreme GPs are found at the end of the distribution, illustrated by a gray area in the BPT diagram in Figure~\ref{fig:BPT}. As~\citet{2016ApJ...826..194S} have already noted, Knots A and B occupy the same region as GPs in the BPT diagram. The figure further shows that both \mk\ and Knot A are fully consistent with the extreme GP region. 
For the integrated \ngc\ system, we estimate a ratio of [OIII]/\hb$=5.05$ and [NII]/\ha$=0.031$, following the same method described in \S\ref{sec:excitation}.  With these values, \ngc\ is again quite consistent with the extreme GPs in the BPT  diagram (Figure~\ref{fig:BPT}).

\mk\ and the GP starbursts therefore have similar ionization parameters $U$ and presumably similar ionizing sources, massive stars.  The average ${\rm log}(U)$ for the extreme GPs is $\sim-2.5$~\citep{Stasinska2015AA,2016ApJ...833..136J}, while~\citet{2016ApJ...816...40J} estimate $-1.9\pm0.3$ for Knot A. Comparing to the LBAs, Figure~\ref{fig:BPT} shows that this class overlaps with average GPs in the BPT diagram, with some LBAs falling in the extreme GPs region, with line ratios $[\textrm{OIII}]\lambda5007/\textrm{H}\beta$ as high as $\sim5.8$, and ${\rm log}[\textrm{NII}]\lambda6584/\textrm{H}\alpha$ as low as $\sim0.04$. As the figure shows, \mk~is however more extreme in its ionization properties than the most extreme LBAs.    

Further indication of a high ionization parameter is strong C III] $\lambda\lambda1907,1909$ doublet emission, often seen in low-metallicity, strongly star-forming galaxies. The C III] doublet is detected in \mk\ with the {\it IUE}/SWP camera \citep{1984A&AS...57..361R}. From the archive spectra we obtain an equivalent width of EW (C III])$\sim14.5\pm2.0$ \AA.  \citet{2016ApJ...833..136J} compute photoionization models of this line with CLOUDY~\citep{1998PASP..110..761F} and investigate the predicted equivalent width of C III] with age, metallicity and ionization parameter. At the metallicity of \mk, $12+{\rm log}({\rm O}/{\rm H})=7.89\pm0.01$ (\S~\ref{sec:metallicity}), the only model that could explain the data is the BPASS instantaneous burst model, which includes both binary interactions and effects from stellar rotation. The model predicts such a strong equivalent width only for ionization parameter ${\rm log}(U)\gtrsim-2$ and an extremely young age of $1$ Myr. We note that the models do not include stars more massive than $150$ M${}_{\sun}$, and there are indications that such may be present in Knot A (\S\ \ref{sec:burstpop}). Examining the CIII]-emitting region in archival {\it IUE} data, the C III] flux comes from an aperture that contains all of \mk, including Knots A and B. Archival {\it HST}/STIS spectra of Knot B show no C III] emission. If the C III] emission is dominated by Knot A, the $1$ Myr estimate above is fully consistent with several other age indicators for Knot A (\S \ref{sec:burstpop}).    
\begin{figure}
\plotone{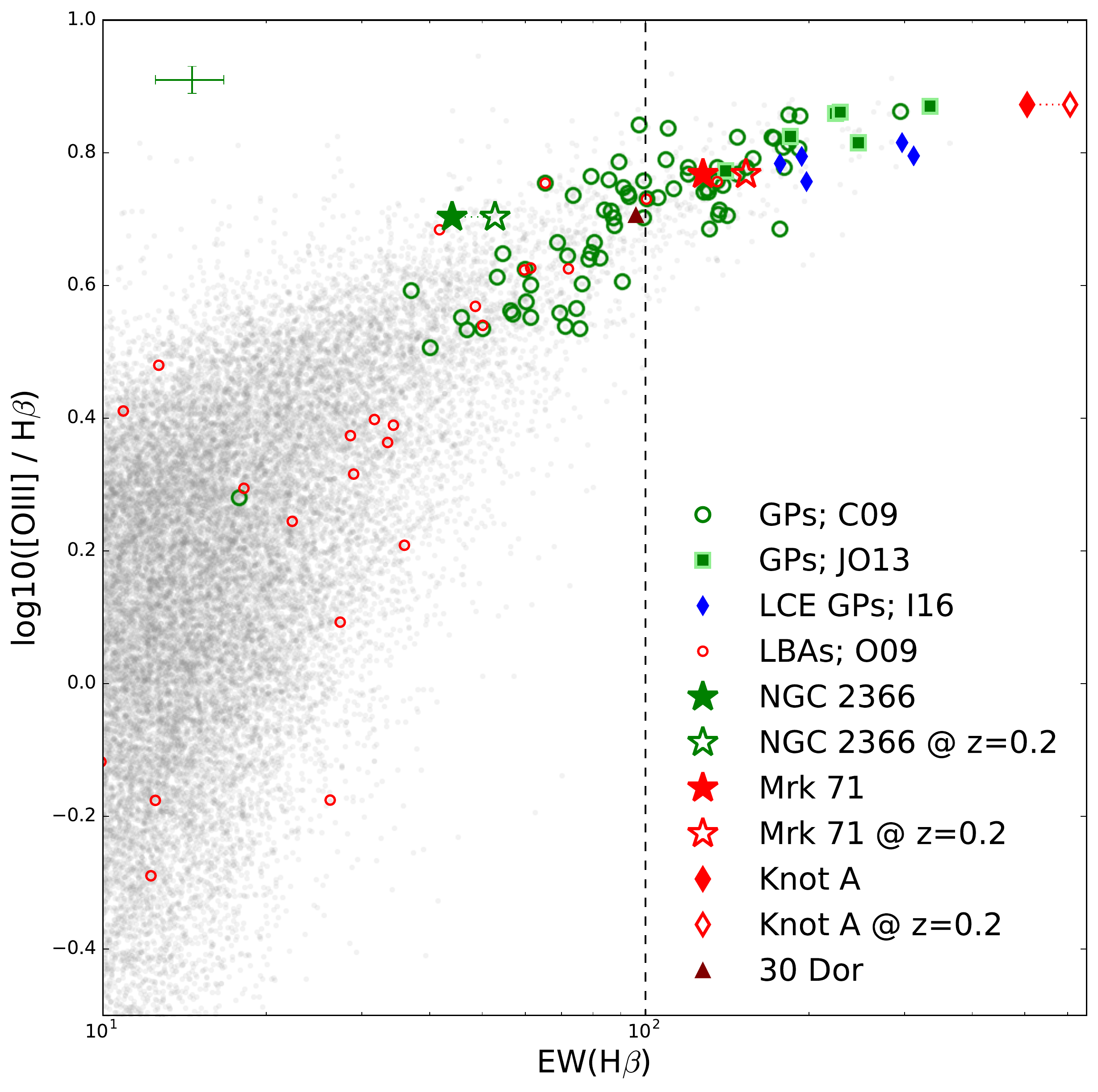}
\caption{The observed equivalent width ${\rm EW}({\rm H}\beta)$ for \mk, its Knot A, and the comparison samples. Extreme GPs, with the confirmed LCE GPs among them, occupy the region with ${\rm EW}({\rm H}\beta)\ge100$\AA\ defined by \citet{Izotov2011}.  Loci for our objects redshifted to GP distances are also shown. \protect\label{fig:ew}}   
\end{figure}

GPs are selected based on their high equivalent width of [OIII]$\lambda5007$. The average GP sample has a mean ${\rm EW}(\rm [OIII])\sim600$ \AA, with the range shown in Table~\ref{tab:comparison}. The equivalent width of \mk~is ${\rm EW}(\textrm{[OIII]})=648.0 \pm 33.0$ \AA~\citep{2006ApJS..164...81M}, which is consistent with this range. Note that the integrated region in the drift scan spectrum of Moustakas \& Kennicutt includes both \mk\ and NGC 2366-II, which dilutes the measured equivalent width of \mk\ alone. For the integrated \ngc\ system, we estimate the total extinction-corrected [OIII]$\lambda5007$ flux by summing the contributions from \mk$+$NGC 2366-II, and NGC 2363 from~\citet{2006ApJS..164...81M}, and the rest of \ngc. For the latter we assume typical SMC \oiii/\hb\ ratios, and convert to \oiii$\lambda5007$ flux by applying the appropriate \ha-based weight from before ($0.14$) to the \hb\ flux.  The continuum flux density is estimated from the integrated $V$-band flux density from~\citet[][$V=10.86$]{2012AJ....144..134H}. The total equivalent width for the \ngc\ system is ${\rm EW}(\rm [OIII])\gtrsim101$ \AA. Note that this is somewhat of an underestimate, since the $V$-band flux contains significant contributions from strong emission lines. The value range for the extreme GPs, obtained from the SDSS DR12, is ${\rm EW}(\textrm{[OIII]})\sim800\mbox{--}2000$ \AA. Thus, the value for the NGC 2366 system is not within the range for the extreme GP sample, but is within that for the average GP sample (Table~\ref{tab:comparison}). EW measurements are sensitive to the star-formation history, and more likely to vary among objects. However, note that our estimate of the equivalent width for Knot A alone is ${\rm EW}(\rm [OIII])\sim2200\pm300$ \AA, based on the [OIII]$\lambda5007$ flux reported by \citet{1997ApJS..108....1I}, and estimating the continuum at $\lambda5007$ from the presented spectrum.

\floattable
\begin{deluxetable}{l|CC|CCCC}
\tablenum{1}
\tablecaption{\mk/\ngc\ as a GP analog\label{tab:comparison}}
\tablewidth{0pt}
\tablehead{
\colhead{} & \colhead{Average GPs} & \colhead{Extreme GPs}& \colhead{NGC 2366} & \colhead{\mk} &  \colhead{Knot A} &  \colhead{Knot B}
}
\startdata
$r_{1/2}$ [kpc] [optical]    & \sim1.0^{n}  & 0.3\mbox{--}2.4^{a,b} & 2.7\pm0.1^{c}   &\nodata &\nodata & \nodata\\
$\textrm{EW}([\textrm{OIII}]\lambda5007)$ [\AA] & 13\mbox{--}2400^{d}  &   800\mbox{--}2000^{a,b}  & >101 &  648.0 \pm 33.0^{e}  &2243\pm345^{f}  &435\pm62^{f}\\
$\textrm{EW}({\rm H}\beta)$  [\AA] & 9\mbox{--}295^{d} & 100\sim300^{a,b} & \sim44.0 &127.6 \pm 5.5^{e}& 505\pm51^{g}&176\pm18^{g} \\
$[\textrm{OIII}]\lambda\lambda5007,4959/[\textrm{OII}]\lambda3727$ &
0.5\mbox{--}14^{\ddagger} & $6 - 14$ ^{a,b}&6.34 & 11.7^{h} &23.0\pm0.6^{i}&9.4\pm0.2^{i} \\
He II $\lambda4686/{\rm H}\beta$ &  0.0086\mbox{--}0.0131^{j}& 0.008\mbox{--}0.02^{a} &\lesssim0.019 &  0.026^{h} & 0.009\pm0.001^{f,h,i}&0.022\pm0.001^{i,f,h} \\ 
$[\textrm{OIII}]\lambda5007/{\rm H}\beta$ &  5.18^{j}& 5.7\mbox{--}7.3^{a,b}  &5.10 & 5.86\pm0.33^{e} & 7.46\pm0.06^{i,h} &6.29\pm0.06^{i,h} \\
$[\textrm{NII}]\lambda6584/{\rm H}\alpha$ &  0.063^{j}& 0.016\mbox{--}0.047^{a,b} &0.031 &0.023\pm 0.020^{e} & 0.0119\pm0.0004^{i}&0.0111\pm0.0004^{i} \\
${\rm log}(U)$  & -3.0~{\rm to}-1.0^{r}  &-2.5^{r}    &\nodata &-2.20\pm0.39^{h} & -1.89\pm0.3^{h} &-2.12\pm0.23^{h}\\
$T_e(\textrm{OIII})$  [K] &\sim13000^{k}&\sim13400\sim15500^{a,b} &\nodata  &\sim15000 &\sim16000^{f,g,i} & \sim14000^{f,g,i}  \\
$n_e(\textrm{SII})$   [cm${}^{-3}$] &\sim180^{k}&\sim100\sim1000^{a,b} &\nodata &\sim200 & 235\pm41^{i} & 163\pm39^{i}   \\
$12+{\rm log}({\rm O}/{\rm H})$& 7.82\mbox{--}8.54^{k,l}  &7.76\mbox{--}8.04^{a,b} & 7.89\pm0.01^{\dagger} & 7.89\pm0.01^{f,m} & 7.89\pm0.01^{\dagger}  & 7.89\pm0.01^{\dagger} \\
$C({\rm H}\beta)$ &  0.02\mbox{--}0.77^{j} & 0.07-0.16^{\dagger,b} &\nodata& 0.13\pm0.04^{h} & 0.2\mbox{--}0.35^{h} & 0.2^{h,i} \\
High-velocity gas [km s${}^{-1}$] &  \textrm{FWZI}\ge1000\mbox{--1750}^{n} & \textrm{FWZI}\sim600^{n}\mbox{--}2460&\nodata & \textrm{FWZI}\sim7000^{o} & \textrm{FWZI}\sim7000^{\dagger} &\textrm{FWZI}\sim7000^{\dagger}  \\
LIS $v_{\rm off}$ [km s$^{-1}$] &-200^{s}&-200^{s}&\nodata&\nodata&\nodata &-154\pm29^{t}\\
He I $\lambda3819/{\rm H}\beta$&\nodata  &\sim0.014^{a}   & \nodata &\sim0.010 & \sim0.012\pm0.001   & 0.007\pm0.005\\
Burst age [Myr]  & 3\mbox{--}5^{k}  &\lesssim4.4^{a,b} &\nodata &\nodata & \lesssim2^{p} & 3\mbox{--}5^{p}     \\
$\textrm{sSFR}({\rm H}\alpha)$ [$10^{-9}\textrm{yr}^{-1}$]    &  1\mbox{--}140^{k} & 7-230^{a,b} &  0.5^{q,c}&\nodata &\nodata&\nodata \\
log(M${}_\star$[${\rm M}_\sun$]) & 7.8\mbox{--}10.0^{k}&8.2\mbox{--}9.6^{a,b}& 8.41^{+0.05}_{-0.10}{}^{q} & \gtrsim5.2  & \sim5.1 & \sim4.3\\
\enddata
\tablecomments{All values are corrected for Galactic and internal reddening.\\
  References, with apertures given for nebular data:\\
  a - \citet{2013ApJ...766...91J}; \\
  b -\citet{2016Natur.529..178I,2016MNRAS.461.3683I}; \\
  c - \citet{2001ApJ...556..773H}; \\
  d - \citet{2009MNRAS.399.1191C}; \\
  e - \citet{2006ApJS..164...81M}; includes \mk\ and NGC
  2366-II, aperture $30\arcsec\times90\arcsec$; \\
  f - \citet{1997ApJS..108....1I}, slit width $2.0\arcsec$; \\
  g - \citet{2016ApJ...826..194S}, slit width $1.3\arcsec$; \\
  h - \citet{2016ApJ...816...40J}, circular aperture of $r=1.8\arcsec$; \\
  i - \citet{1994ApJ...437..239G}, slit width $1.2\arcsec$; \\
  j - \citet{2012PASP..124...21H}; \\
  k - \citet{Izotov2011}; \\
  l - \citet{2010ApJ...715L.128A}; \\
  m - \citet{1999AJ....117.2789H}; \\
  n - \citet{2012ApJ...754L..22A}; \\
  o - \citet{2009AA...500..817B}; \\
  p - \citet{2000AJ....119..688D}; \\
  q - \citet{2014AA...566A..71L}; \\
  r - \citet{Stasinska2015AA}, \citet{2016ApJ...833..136J}; \\
  s - \citet{2015ApJ...809...19H};\\
  t - \citet{2011AJ....141...37L};  \\
  u - \citet{2010ApJ...721..297M};\\
  $\ddagger$ - Obtained from SDSS DR12 and corrected for internal extinction assuming $C({\rm H}\beta)=0.2$ and
SMC dust law; \\
  $\dagger$ - Value applies to multiple components, and is adopted 
from the column with references. }
\end{deluxetable}

\subsection{Specific Star Formation Rate}
Both average and extreme GPs are characterized by high sSFR with
ranges sSFR$({\rm H}\alpha)=10^{-7}$--$10^{-9}$
yr${}^{-1}$ by \citet{Izotov2011}. They obtained the current SFR from the H$\alpha$ luminosities of the GPs, while the stellar mass estimates were derived by approximating the star formation history (SFH) by a short young burst and an older stellar population, and fitting the spectral energy distribution of each GP. 
While the typical GP appears as a compact object (\S~\ref{sec:morph}),
some GPs show evidence of extended underlying stellar populations.
Given the distances and low surface brightness, we caution that an
underlying faint extended population may not be included in current
stellar mass estimates from the literature. The sSFR for the GPs
therefore should be considered upper limits.

The current SFR of the host galaxy NGC 2366 is $0.13\pm0.03$
M${}_{\sun}$ yr${}^{-1}$, derived from the H$\alpha$ luminosity within
a $2.6$\arcmin\ aperture~\citep{2001ApJ...556..773H}. This value is
consistent with estimates from modeling the observed color-magnitude
diagram of NGC 2366~\citep[Figure 9 in][]{2010ApJ...721..297M}. The
stellar mass of NGC 2366 was reported by~\citet{2014AA...566A..71L},
who integrate over the SFH in McQuinn et al. and obtain a stellar mass
of $26\pm 3\times 10^{7}$ M${}_{\sun}$. For NGC 2366 one therefore
obtains sSFR$=5\pm 1 \times 10^{-10}$ yr${}^{-1}$. This is a
factor of two lower than the lowest sSFR observed for the average GP
sample. The young population in NGC 2366 is dominated by \mk~\citep{2005ApJ...627..739T},  and we estimate its stellar mass at $M_\star\sim 1.6\times10^{5}$ M${}_{\sun}$ (\S\S \ref{sec:burstpop}, \ref{sec:clusterB}). Stellar masses of the young population in GPs are not widely available, but note that the LCE GP in~\citet[][]{2016Natur.529..178I} has an estimated burst stellar mass of $M_\star=2.4\pm0.3\times10^{8}$ M${}_{\sun}$, which is orders of magnitude larger. Our GP analog therefore samples the low-mass faint end of the GP distribution.

If we consider \mk\ as a dominant compact object in NGC 2366, the system is similar in morphology to LBAs (\S~\ref{sec:morph}). Being more massive, LBAs have a slightly lower range of sSFR$({\rm H}\alpha)=10^{-8}$--$10^{-10}$ yr${}^{-1}$~\citep{2009ApJ...706..203O}, which is more consistent with the value for NGC 2366.   

A consequence of high sSFR is a large equivalent width of
H$\beta$. In fact, \citet{Izotov2011} suggest using the observed H$\beta$
equivalent width of $\ge100$\AA\ as a selection criterion for GP-like
galaxies with strong star formation. We show ${\rm EW}({\rm H}\beta)$
for the GP and LBA samples in Figure~\ref{fig:ew}, and note that the
Izotov et al. criterion of observed EW$\ge100$\AA\ includes about half of the
average GP sample and the full sample of extreme GPs, while the only
LBA which makes the cut is also a GP. For the average GPs, the range in
${\rm EW}({\rm H}\beta)$ in Table~\ref{tab:comparison} is obtained from the SDSS DR12.  For the extreme GPs, the range of ${\rm EW}({\rm H}\beta)$ is $100$ to $\sim300$\AA~\citep{Izotov2011}.
In Figure~\ref{fig:ew} and Table~\ref{tab:comparison} we show our estimate of the equivalent width of \hb\ for the host galaxy \ngc.  We obtain this value by assuming the total \hb\ luminosity is $2.86$ times less than the total \ha\ luminosity for \ngc~\citep[$L({\rm H}\alpha)=1.3\times10^{40}$ erg s${}^{-1}$;][]{2008ApJS..178..247K}, using a distance of $3.4$ Mpc, and estimating the continuum emission from the integrated $B$-band flux density in~\citet[][$B=11.15$]{2001ApJ...556..773H}. The ${\rm EW}({\rm H}\beta)\sim44.0$\AA\ for \ngc, which is consistent with average GPs values. 

Drift-scan spectra of \mk\ from~\citet{2006ApJS..164...81M}, yield ${\rm EW}({\rm H}\beta)=127.6 \pm 5.5$\AA.  These data include a contribution from NGC 2366-II, as discussed in Sec.~\ref{sec:excitation}.  For Knot A, which is responsible for most of the similarities between
\mk\ and GPs, ${\rm EW}({\rm H}\beta)=505 \pm 51$ \AA\ within a $1.3$\arcsec\ aperture~\citep{2016ApJ...826..194S}. When isolated, it thus
has a more extreme equivalent width than even the extreme GPs, as shown in Figure~\ref{fig:ew}. 

\subsection{$T_{e}$ and $n_e$}\protect\label{sec:TeNe}
In extreme GPs the $[\textrm{OIII}]\lambda\lambda5007,4959/\lambda4363$ ratio indicates electron temperatures $T_{e}\sim15000$ K, while from $[\textrm{SII}]\lambda6716/\lambda6731$ one obtains electron densities $n_e=100$--$1000$ cm${}^{-3}$~\citep{2013ApJ...766...91J}. The electron temperature has been measured in the nebular emission from Knot A by several authors~\citep{1994ApJ...437..239G,1997ApJS..108....1I,2016ApJ...826..194S} and is on average $\sim16000$ K.  \citet{1994ApJ...437..239G} also measure the electron density of the Knot A to be $n_e=235\pm41$ cm${}^{-3}$.  We note that most of the nebular measurements are generally based on apertures on the order of 1\arcsec\ and are therefore characteristic of the SSC environment, rather than conditions within Knot A itself.  The reported values are consistent with electron temperatures and electron densities seen in the extreme GP sample (Table~\ref{tab:comparison}.) 

\subsection{Abundances}\protect\label{sec:metallicity}
 
Average GPs have sub-solar abundances with a range $12 + {\rm log}({\rm O}/{\rm H}) \sim 7.8 - 8.5$, based on the so-called ``direct'' method, which uses the electron temperature derived as above, from the [O III] $\lambda\lambda4959,5007/\lambda4363$ ratio to obtain the oxygen abundance of the ionized gas from the collisionally excited lines of [OIII] and [OII] \citep[e.g.,][]{2010ApJ...715L.128A,Izotov2011}. The extreme GPs have a range spanning lower abundances, $12 + {\rm log}({\rm O}/{\rm H}) = 7.8 - 8.0$~\citep{Izotov2011,2016Natur.529..178I,2016MNRAS.461.3683I}. The abundance of \mk\ as measured from Knot~A is $12 + {\rm log}({\rm O}/{\rm H}) = 7.89\pm0.01$, using the same method~\citep{1997ApJS..108....1I}, and is thus consistent with both average and extreme GPs.  

Another method for obtaining abundances  uses optical recombination
lines like O II$\lambda4649$.  Although these lines are often too weak to be
detected, the derived abundances are much more robust, and almost independent of
temperature.  Using recombination lines to obtain the oxygen abundance for
\mk,~\citet{2002ApJ...581..241E} find a higher value of $12 + {\rm
  log}({\rm O}/{\rm H}) = 8.19\pm0.11$. This is in good agreement
with~\citet{1999ApJ...527..110L} who are only able to reconcile the
observed emission-line spectrum of \mk\ with photoionization
models when using a higher abundance of $12 + {\rm log}({\rm O}/{\rm H}) = 8.2$.  This higher abundance for \mk\ would imply more dust, which is to the detriment of \lyc~escape since dust readily attenuates
ionizing radiation. However, the confirmed LBA \lyc-leaker found
by~\citet{2014Sci...346..216B} has an abundance of $12+{\rm log}({\rm
  O}/{\rm H})=8.67$, which is much higher than the abundance in
\mk. Similarly high abundances are found for all three LBAs which are
LCE candidates in~\citet{2011ApJ...730....5H}.  We caution that the LBA
abundances were estimated using the semi-empirical ``O3N2'' method
from~\citet{2004MNRAS.348L..59P}, but even if there is a systematic
overestimate, the obtained values suggest that the LBA abundances are
in any case higher than for GPs.

We stress that, regardless, the abundance of \mk~is fully consistent with those of both average and extreme GPs when consistently determined using the same ``direct'' method, as described above.

\subsection{Reddening and Dust}\protect\label{sec:dust}
The average GPs have a low extinction correction factor $C({\rm H}\beta)$ ranging between $0.02\mbox{--}0.77$~\citep{2012PASP..124...21H}, with the extreme LCE GPs on the lower end of this distribution with values $0.07\mbox{--}0.13$~\citep{2016Natur.529..178I,2016MNRAS.461.3683I}. For the extreme GPs in~\citet{2013ApJ...766...91J} we measure $C({\rm H}\beta)\leq0.16$ from SDSS spectra, and this is the maximum value we present in Table~\ref{tab:comparison} for the combined extreme GP sample. The average $C({\rm H}\beta)$ in \mk~is $\sim0.13\pm0.04$~\citep{2016ApJ...816...40J}, consistent even with the extreme GP sample. The low average $C({\rm H}\beta)$ factor implies a low dust content. 

The presence of dust can also be inferred from emission in the IR regime.~\citet{2011A&A...536L...7I,2014A&A...561A..33I} study the IR emission from hot dust of a sample of compact star-forming $z<0.6$ galaxies using data from {\it WISE}. Among these are $16$ GPs from the average GP sample. The $3.4\mu{\rm m}-4.6\mu{\rm m}$ and $4.6\mu{\rm m}-22.0\mu{\rm m}$ colors are a proxy of the slope of the spectrum at these wavelengths. For the GPs the observed average colors are $1.9$ and $7.3$, respectively, indicating the presence of hot dust.  However, we stress that these GPs are not representative of the average GP sample since they were selected on the basis of having very red WISE colors. 

To compare to \ngc, we use the spatially integrated IR values
available as part of the Spitzer Local Volume Legacy 
survey in IRAC and MIPS bands~\citep{2009ApJ...703..517D}. To cover
the same wavelength range we use IRAC1 ($3.6\mu{\rm m}$), IRAC2 ($4.5\mu{\rm m}$), and MIPS1 ($24\mu{\rm m}$), and obtain $3.6\mu{\rm m}-4.5\mu{\rm m}=-0.3\pm0.2$, and $4.5\mu{\rm m}-24\mu{\rm m}=0.3\pm0.2$. These colors are typical of the star-forming galaxies in~\citet{2009ApJ...703..517D}. While the filters are slightly different, it cannot account for the large differences with the WISE observations. Due to the selection effect of the WISE GPs we can only conclude that there exist GPs which have much steeper slopes over the wavelength range $3.6\mu{\rm m}\mbox{--}24\mu{\rm m}$, and therefore likely more dust than \ngc.

 \subsection{Column density of neutral gas}\protect\label{sec:coldens}
Resonant UV absorption lines produced in the neutral interstellar
medium (ISM) of star-forming galaxies are sensitive to its opacity and
covering factor~\citep[e.g.][]{2001ApJ...558...56H,2011ApJ...730....5H,2011AJ....141...37L}. In
LBAs, \citet{2011ApJ...730....5H} use the low equivalent widths of
interstellar C II $\lambda 1335$ and Si II $\lambda1260$ absorption
lines to infer significant leakage of \lyc~from three out of eight
LBAs in their sample. In extreme GPs the non-detection of these
absorption lines has been suggested as an indication of very low
column density of the intervening gas, increasing the likelihood of
\lyc~leakage from these objects~\citep{2014ApJ...791L..19J}. 

Even if such absorption is detected one may still be in the linear
part of the curve of growth, where the column density is low. Using
HST/FOS and GHRS observations of local starbursting galaxies,
\citet{2011AJ....141...37L}, show that the equivalent widths of
several low-ionization state, UV absorption lines (LIS) like Si II $\lambda1260$,
OI$-$Si II $\lambda1303$, Si II $\lambda1526$, and C II $\lambda1335$
for \mk\ are among the weakest in their sample of local starburst
galaxies, while Fe II at $\lambda1608$, $\lambda2370$, $\lambda2600$,
and Mg II $\lambda2800$ are not even detected.  This suggests that the
line of sight towards \mk\ may be optically thin, although this could
also be due to its low metallicity.  However, we note that significant
absorption in these lines is detected in I Zw 18 and SBS
0335--052~\citep{2014ApJ...795..109J}, which have metallicities much
lower than that of \mk.

Using the ratio of Si II $\lambda1260$/$\lambda1526$, \citet{2011AJ....141...37L} demonstrate that most of their sample of local starburst galaxies is optically thick in Si II, with $\lambda1260/\lambda1526\sim0.5$--$2$. However, for Knot A in \mk, their data show that the corresponding ratio is $\lambda1260/\lambda1526=6.0\pm0.3$ \citep[][their Table 8]{2011AJ....141...37L}, where optically thin values correspond to a ratio of $\sim5.2$.  Thus, the Si II transitions are optically thin, implying similar conditions for \hi. 

Under optically thin conditions, we can obtain the column density of Si II from the equivalent width of, e.g., the Si II$\lambda1260$ line, from which we obtain $N({\rm Si\ II})=2\times10^{14}$ cm$^{-2}$. To translate this to \hi\ column density one would need the Si II abundance, which in turn depends on the total Si abundance and the Si ionization structure inside of the cloud. Modeling the latter is beyond the scope of this paper, and so we provide only a coarse upper limit by approximating that \hi\ coexists exactly with Si II. The total silicon abundance for \mk\ is $12+\log\rm(Si/H)=6.3$~\citep{Garnett1995}, giving an upper limit on the \hi\ column density of $N({\rm H\ I})\lesssim10^{20.0}$ cm${}^{-2}$. While consistent with optically thin Si II transitions, this upper limit is too coarse to definitively demonstrate optically thin conditions for \hi, which occur for $N({\rm H\ I})\lesssim10^{17.2}$ cm${}^{-2}$.

Another good tracer of neutral, diffuse gas is absorption in the
resonance doublet Na I $\lambda\lambda5890,5896$ in the optical. In
\mk, \citet{2004ApJ...610..201S} do not detect Na I in absorption and
put upper limits on the column density of Na I at $<0.57\times10^{12}$
cm${}^{-2}$. They suggest this to be indicative of a lack of neutral gas
in the line of sight, and significant leakage of ionizing radiation
from this galaxy. Similarly to Si II, we can apply the same method to estimate an \hi\ column density from the upper limit on the Na I doublet. Taking the SMC abundances as representative of \mk\ (\S \ref{sec:metallicity}), $\log\rm(Na/H)\sim -8.50$
\citep{2007A&A...470..941C}. Approximating that \hi\ coexists exactly
with Na I, the upper limit on the Na I non-detection implies
an H I column density of $N({\rm H\ I})\lesssim10^{20.5}$
cm${}^{-2}$. This upper limit is again not a sufficient condition for \lyc~escape. Taking the abundance of Na I to be equal with that of total Na is even less likely to be a good approximation here since the first ionization potential of Na is only $5.1$ eV, and thus much of the Na could be ionized in an otherwise neutral hydrogen cloud. But in any case, the non-detection in absorption of Na I
and other low-ionization species, as well as the optically thin Si II transitions, are consistent with \lyc~leakage from \mk. 

\subsection{Kinematics}\protect\label{sec:kin}
\citet{2012ApJ...754L..22A} select five GPs from the~\citet{2009MNRAS.399.1191C} sample for high-resolution follow-up spectroscopic observations, and detect broad ${\rm H}\alpha$ wings of full width zero intensity ${\rm FWZI}>1000$ km s${}^{-1}$ in all of them.  We note that one of these objects, SDSS J004054.31+153409.8, falls into the extreme GP region in Figure~\ref{fig:BPT} with log$([\textrm{OIII}]/\textrm{H}\beta)=0.8$ and log$([\textrm{NII}]/\textrm{H}\alpha)=-1.4$, while the other four have more typical excitation, with average log$([\textrm{OIII}]/\textrm{H}\beta)=0.6$ and log$([\textrm{NII}]/\textrm{H}\alpha)=-1.0$. For the extreme GPs in~\citet{2013ApJ...766...91J}, we measure a range of ${\rm FWZI}=770\mbox{--}2460$ km s${}^{-1}$. Since the broad wings are found in both the average and the extreme GP samples, the kinematics may have similar origin. Amor\'in et al. interpret these features as rapid outflows of ionized gas due to strong stellar winds from massive stars in combination with expansion of multiple supernova remnants. At the redshift of these GPs, $z\sim0.2$, their unresolved appearance prohibits any detailed testing of this scenario. However, similarly broad features are present in \mk, which might provide clues to their origin in GPs.  In \mk\ a faint, broad spectral component of full width half maximum ${\rm FWHM}\sim2400$ km s${}^{-1}$ (FWZI$\sim7000$ km s${}^{-1}$) in ${\rm H}\alpha$ and [O III]$\lambda5007$ was detected by \citet{1992ApJ...386..498R} and \citet{1994ApJ...437..239G} over most of the complex.  These authors found that conventional mechanical feedback from stellar winds and supernovae are unlikely to explain the high velocities. Recently,~\citet{2009AA...500..817B} showed that the observations can be reproduced by turbulent mixing layers, in which the broad wings result from acceleration of photoionized turbulent gas entrained from dense clumps by a strongly supersonic SSC wind. 

Outflows of interstellar gas accelerated by massive stellar winds and
supernovae can be traced by UV absorption lines that are blueshifted
with respect to the systemic velocity. Such blueshifting is detected
in, for example, the C II $\lambda1335$ and Si III $\lambda1206$ lines
in LBAs, implying strong outflow velocities of gas reaching $\sim1500$
km s${}^{-1}$~\citep{2011ApJ...730....5H}.  We caution that in the
case of LCE LBAs the outflow may be AGN-assisted since these galaxies
are found in the SF-AGN composite region in the BPT diagram; however
no evidence of any AGN has been found~\citep{2009ApJ...706..203O}.~\citet{2014ApJ...791L..19J} and~\citet{2015ApJ...809...19H} find broad and blueshifted Si II $\lambda1260$ and C II $\lambda1335$ absorption lines for two extreme GPs, indicative of outflow of cool neutral gas. \citet{2011AJ....141...37L} observe these latter two lines in \mk, which also appear offset from photospheric features for Knot~B. The velocity offsets $v_{\rm off}$ are on the order of $200$ km s$^{-1}$ in both GPs and Mrk 71, as shown in Table~\ref{tab:comparison}. We caution that the value for Knot~A published by \citet{2011AJ....141...37L} is unreliable since there are no known photospheric lines observed in this object, to serve as the systemic reference \citep{2000AJ....119..688D}.

\subsection{Stellar population of Knot A}\protect\label{sec:burstpop}
Knot A hosts a massive, enshrouded SSC.  As mentioned earlier, no stellar features have ever been confirmed in this object, and so we rely on indirect inference of its stellar properties. Using a total \ha\ luminosity for \mk\ of $L({\rm H}\alpha)=1.4\times10^{40}$ erg s${}^{-1}$, which is dominated by Knot A (\S\ \ref{sec:bigGP}),~\citet{1994ApJ...437..239G} estimate a lower limit to the total stellar mass of M${}_\star\gtrsim3.4\times10^{5}$ M${}_{\sun}$. This value differs by a factor of $\sim6$ from the estimate of M${}_\star= 5.3\times10^{4}$ M${}_{\sun}$ for Knots A and B combined, by~\citet{2016ApJ...826..194S}, who instead normalize by the $V$ band luminosity. This difference in stellar mass estimates can likely be reconciled by considering the extinction. Sokal et al. assume an extinction $A_V=0.0$ for Knot A, which is an understimate since the SSC is still enshrouded in its natal cloud (\S\ref{sec:bigGP}). While much of the literature reports $C({\rm H}\beta)\approx0.2$ for Knot A, this is usually obtained from ground-based apertures on the order of 1\arcsec, and  it is clear from the $C$(\hb) map from~\citet{2016ApJ...816...40J}, which is based on new high-resolution HST data, that a lower limit to the extinction correction is $C({\rm H}\beta)\gtrsim0.35$ for the dense knot itself. A factor of $6$ increase to the $V$ band flux requires $A_V=1.9$, which translates to $C({\rm H}\beta)=0.68$ using the~\citet[][$R_V=4.1$]{2000ApJ...533..682C} dust attenuation law.  A comparable extinction of $A_V\geq1.5$ is typical among very young clusters with ages $\leq3$ Myr~\citep[][]{2002AJ....124.1418W}. There are also additional differences between the methods used to obtain the mass estimates.~\citet{1994ApJ...437..239G} use solar metallicity evolutionary models and mass loss prescription, and an assumed age of $3\mbox{--}5$ Myr; while~\citet{2016ApJ...826..194S} use low-metallicity, high-mass loss models.  Accounting for these differences in inputs can further help to reconcile the two mass estimates.  

We perform our own estimate of the stellar mass for Knot A by scaling with the revised \ha\ luminosity measurement for \mk\ by~\citet[][]{2016ApJ...816...40J}, $L({\rm H}\alpha)=8.4\times10^{39}$ erg s${}^{-1}$. We use a ``standard'' {\sc Starburst99} model~\citep[v7.0;][]{2014ApJS..212...14L}, adopting a Kroupa IMF, instantaneous starburst, and $Z=0.2Z\sun$. We assume that Knot A has an age of $1$ Myr and \ha\ luminosity that is $90\%$ that of \mk\ (\S\ref{sec:bigGP}). The resulting stellar mass of Knot A is $\sim1.3\mbox{--}1.4\times10^{5}$ M${}_{\sun}$, depending on whether we use tracks without or with rotation, respectively. This is our adopted stellar mass for Knot A that we list in Table~\ref{tab:comparison}.  

The high ionization parameters in extreme GPs,  ${\rm log}(U)=-2.5$, suggest very young ages. Stellar population synthesis models, together with photoionization models, suggest an upper age limit of $\lesssim4$ Myr~\citep{2013ApJ...766...91J}.  Knot A in \mk~has a comparable ionization parameter ${\rm log}(U)=-1.89 \pm 0.3$~\citep[][]{2016ApJ...816...40J} and excitation as the GPs (Figure~\ref{fig:BPT}), and therefore could be similar in age. The GP ages are consistent with a paradigm in which mechanical feedback has enough time to facilitate \lyc~escape by punching holes in the ISM, while hot, massive stars of ages $3$--$5$ Myr, dominated by WR stars, produce the escaping ionizing photons \citep[e.g.,][]{2011ApJ...731...20F,2013ApJ...779...76Z}. 

There are indications, however, of an even younger age for Knot A,
which could mean a possible shift in this paradigm.  Here again, the
properties of Knot~A bear strong resemblance to those of extreme
GPs. In particular, the presence of nebular He I $\lambda 3819$ is
readily detected in the extreme GPs \citep{2013ApJ...766...91J}, with
average He I$ \lambda3819 / {\rm H}{\beta}=0.014\pm0.006$ and EW(He I
$\lambda 3819$)$=1.95$ \AA. The He I $\lambda 3819$ line is also
clearly detected in Knot A, as seen in the spectrum of
\citet{2016ApJ...826..194S}. Applying the extinction correction for
Knot A, $C$(\hb)$\gtrsim0.35$ (\S\ \ref{sec:dust}), we measure a line
ratio and equivalent width in this spectrum similar to GPs: ${\rm He~I
  \lambda3819 /H}\beta=0.012\pm0.001$ and ${\rm EW}(\textrm{He I}
\lambda3819)=1.9\pm0.3$\AA. Using evolutionary synthesis models,
\citet{1999ApJS..125..489G} show that this line is strongest at ages
$0$--$2$ Myr, when EW(He I $\lambda 3819$)$=2.6$--$1.5$ \AA,
respectively, and significantly weakens or disappears at $\ge3$ Myr
since it becomes dominated by stellar absorption.  Since 
Knot A is dominated by nebular emission and no stellar features are
visible, the interpretation of this nebular line is more ambiguous.
However, its presence suggests that both the extreme GPs and the starburst in
Knot A are of similarly young age $\lesssim2$ Myr.  This is consistent
with an age estimate for Knot A of $\lesssim 1$ Myr by
\citet{2000AJ....119..688D}, based on the still strongly embedded
condition of the young super star cluster, also stressed by \citet{2016ApJ...826..194S}.  

Another indication of Knot A's very young age is the lack of classical
WR stars~\citep{2000AJ....119..688D}, which appear at ages around 3
Myr.  They can be identified by a broad ``blue bump'' in the continuum
near $\lambda4650$, due to N III $\lambda\lambda4634\mbox{--}4641$ and
He II $\lambda4686$ emission associated with WN stars, indicating an
age of $\sim3 - 5$ Myr.  While there is narrow, nebular He II
$\lambda4686$ emission clearly detected in the~\citet{2009AA...500..817B} echelle spectrum of Knot A in \mk,
there also seems to be a very faint, broad bump centered only at
$\lambda$4686.  Although those authors attribute this faint feature to
turbulent mixing layers, it could also be a faint detection of WN
features from very massive stars~\citep[VMS,
  e.g.,][]{2010MNRAS.408..731C,2011MNRAS.416.1311C,2015A&A...578L...2G,2016ApJ...823...38S}. VMS
are $150\mbox{--}300$ M${}_{\sun}$ O-type supergiants, which have
short lifetimes of $1\mbox{--}3$ Myr and are found in high-mass
($\gtrsim 10^{4}$ M${}_{\sun}$), extremely young
clusters~\citep[e.g.][]{2010MNRAS.408..731C}. The VMS are
of class O2--3.5 If${}^{*}$ at the zero age main sequence, and within
the first $1\mbox{--}2$ Myr of their lives become ``slash'' stars of
class O2--O3.5 If${}^{*}$/WN5--7, with WR features in their spectra  
\citep{2011MNRAS.416.1311C}. Such VMS stars of initial masses up to
320 M$_\sun$ have been identified in R136 in the 30 Doradus star-forming region of the Large Magellanic Cloud~\citep{2010MNRAS.408..731C}. Additionally,~\citet{2016ApJ...823...38S} show that in cluster $\#5$ in NGC 5253, the broad He II $\lambda4686$ emission likely arises from stellar emission of VMS with $\ge100$ M${}_{\sun}$ and ages of $1\mbox{--}2$ Myr.  The presence of VMS stars in Knot~A has already been suggested by~\citet{2016ApJ...816...40J} to account for the extreme stellar temperatures required to ionize \heii. If the faint, broad He II emission in Knot A is real, it could further support the presence of VMS stars, which would be consistent with its implied age of $\lesssim2$ Myr. 

On a final note, we again consider the detection of the C III] $\lambda1909$ doublet in \mk.  As discussed in \S \ref{sec:excitation}, the only model prediction of~\citet{2016ApJ...833..136J} that could explain the observed C III] EW$=14.5\pm2.0$\AA\ in \mk\ is the one with an age of $1$ Myr. The observed equivalent width appears to rule out models for ages $\ge3$ Myr, implying that this emission is more likely to come from Knot A, since Knot B has a well established age of $3\mbox{--}5$ Myr, based on its observed stellar population (\S \ref{sec:clusterB}).


\subsection{Knot B}\protect\label{sec:clusterB}
In the above comparisons, we have concentrated on the properties of Knot
A, since it strongly dominates the ionizing luminosity and excitation
of \mk\ \citep[e.g.,][]{1994ApJ...437..239G,2000AJ....119..688D}. However, while not as extreme in
its properties, the exposed cluster, Knot B, also contributes to the GP-like properties of the complex. 

Knot B is a lower-mass SSC with stellar mass $1.2\times10^{4}$ M${}_{\sun}$ reported by \citet{2016ApJ...826..194S}. We obtain a similar mass of $1.5\times10^{4}$ M${}_{\sun}$ with the same {\sc Starburst99} model as for Knot A, but assuming an age of $3$ Myr and only $11\%$ of the \mk\ $L({\rm H}\alpha)$ (\S\ \ref{sec:bigGP}).  Knot B's stellar population has been spectroscopically studied with ground-based observations from the William Herschel Telescope by \citet{1994ApJ...437..239G} and {\it HST}/FOS observations by \citet{2000AJ....119..688D}. \citet{1994ApJ...437..239G} detect P-Cygni, stellar wind signatures from C IV $\lambda 1550$ and N V $\lambda 1240$ doublets, as well as WR features consisting of both a blue bump at $\lambda 4660$ and a red bump at $\lambda5800$ \AA, indicating the presence of WC stars.~\citet{2000AJ....119..688D} perform UV spectral synthesis of Knot B and conclude that around $800$ B and $40$ O stars must be present, which can account for only $\sim 11$\% of the total number of photons needed to ionize \mk\ (\S \ref{sec:bigGP}).  Through {\it HST} narrowband imaging and long-slit spectroscopy, the existence of broad WR features in emission at $4660$ \AA\ and $5810$ \AA\ has been well established, as well as a strong, narrow He II $\lambda4686$ line of nebular origin \citep{1993AJ....106.1460D,1994ApJ...437..239G,2000AJ....119..688D}. There are three known WR stars in the core of Knot B, with their global spectrum dominated by an early-type WC4 star as indicated by the $\lambda4650/5810$ ratio~\citep{2000AJ....119..688D}. \citet{2016ApJ...826..194S} suggest that as many as 8 WR stars are present. The presence of these classical WR stars sets the age of Knot B at $3\mbox{--}5$ Myr. 

The excitation for the region around Knot B is high, as traced by the [OIII]/[OII] ratio of $\sim9.4\pm0.2$~\citep{1994ApJ...437..239G}. The implied ionization parameter is also high, as found by \citet{2016ApJ...816...40J}, who obtain $\log U=-2.12\pm0.23$. This is consistent with both excitation dominated by WR stars and the likely presence of significant Knot A contribution.~\citet{1994ApJ...437..239G} find an electron temperature and density in Knot B similar to those in Knot A, namely, $T_{\rm e}\sim14000$ K, and $n_{\rm e}=163\pm39$. 

In contrast to Knot A, where the Si II $\lambda1260/\lambda1526$ ratio is optically thin (see \S \ref{sec:kin}), in the line of sight to Knot B, the value of $\lambda1260/\lambda1526=0.4$~\citep{2011AJ....141...37L}, and thus suggests an optically thick line of sight. Information about Knot B is included in Table~\ref{tab:comparison}. 

\begin{figure*}
\plotone{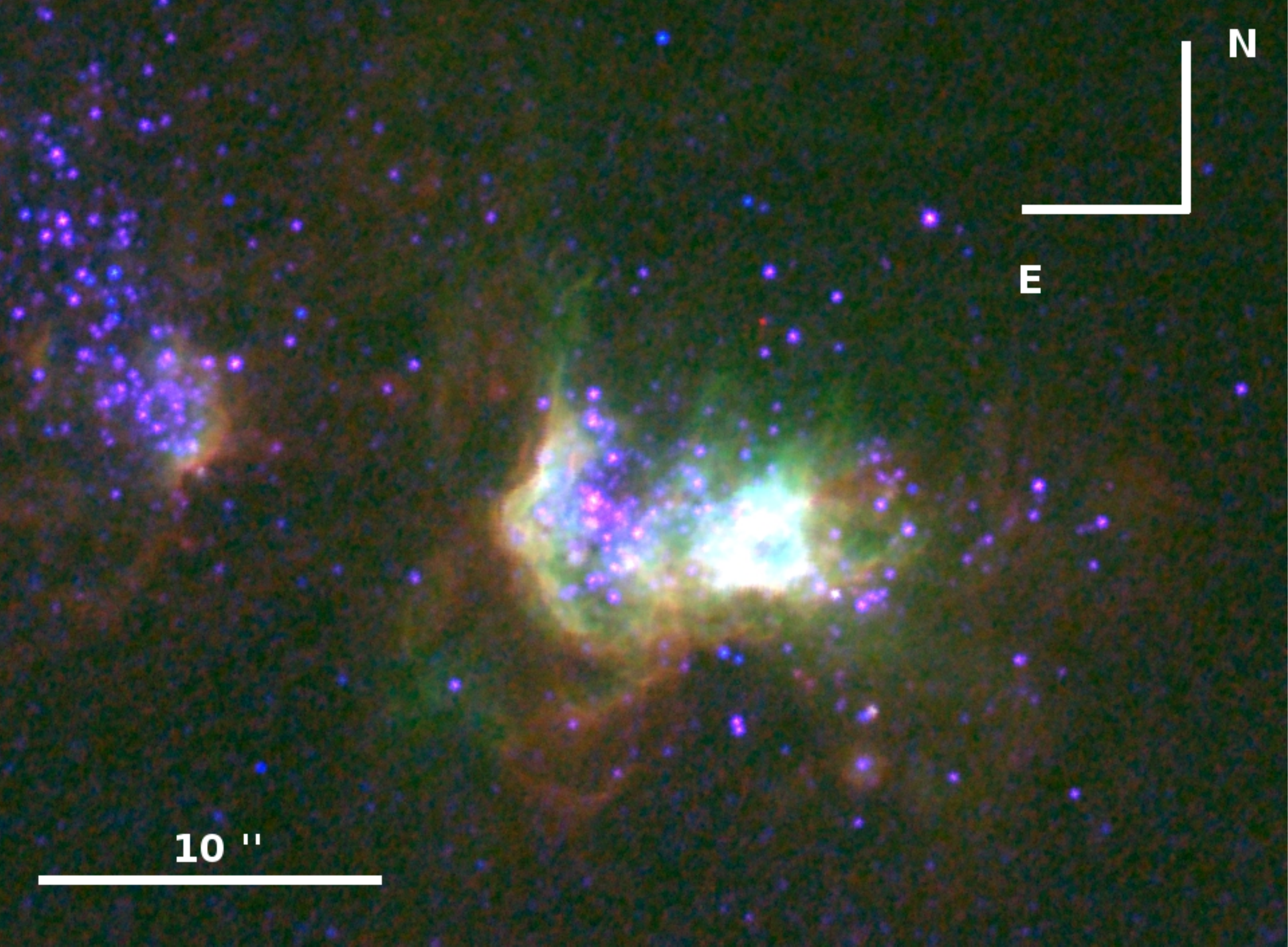}
\caption{Three-color image of \mk, in F373N ([O II]$\lambda3727$;
  red), F502N ([OIII] $\lambda 5007$; green), and F469N (He II
  $\lambda4686$; blue), showing the highly ionized \oiii-dominated
  region extending to the north from \mk.\protect\label{fig:rgb}} 
\end{figure*}

As seen in Figures~\ref{fig:nomenclature} and \ref{fig:rgb}, Knot B
appears to have generated a superbubble with strong shell morphology
to the east, and a blow-out region to the north.  This is consistent
with the substantial mechanical feedback generated by a massive,
somewhat evolved super star cluster. Fabry-Perot observations
by~\citet{1991ApJ...367..141R} confirm that the shell is expanding,
with line splitting in ${\rm H}\alpha$ and [OIII] showing expansion
velocities of $\sim 20$ km s${}^{-1}$. These kinematics are seen in a
region centered on the blowout, which, as seen in Figure~\ref{fig:HI},
extends over a $200$-pc region to the north and coincides with a
low-density chimney seen in VLA HI line observations of NGC 2366 in
the LITTLE THINGS
survey~\citep[e.g.,][]{2008AJ....136.2563W,2012AJ....144..134H}.
The blowout subtends a projected angle of $\sim30^{\circ}$, which
corresponds to a solid angle of $2\%$ out of $4\pi$ steradians,
assuming axisymmetric geometry.  This may be an upper limit
due to projection effects. X-ray
emission from the blowout is marginally detected in {\it XMM-Newton}
observations by \citet{2014MNRAS.441.1841T}, as expected from
mechanical feedback.  

As discussed in Section~\ref{sec:kin}, outflows have been detected in both GPs and
LBAs, and are likely to be accompanied by low column densities.
While the observed velocities seen in GPs and LBAs are much higher, we
note that the blowout seen here appears to be largely transverse to
the line of sight, minimizing the observed kinematics.  It is likely
that mechanical feedback from Knot B cleared large areas in \mk,
reducing the optical depth for Knot A as well.  We discuss this
further in \S \ref{sec:mklyc}.  

\section{\ngc~as LCE candidate}\protect\label{sec:mklyc}

We have established above that the properties of \ngc, and \mk, driven by Knot A in particular, are quantitatively consistent with GPs and spectroscopically confirmed LCE GPs, in terms of morphology, excitation, sSFR, abundances, dust reddening, neutral gas column density and gas kinematics.  We now further examine the likelihood that \ngc\ is an LCE.  

\begin{figure*}
\plotone{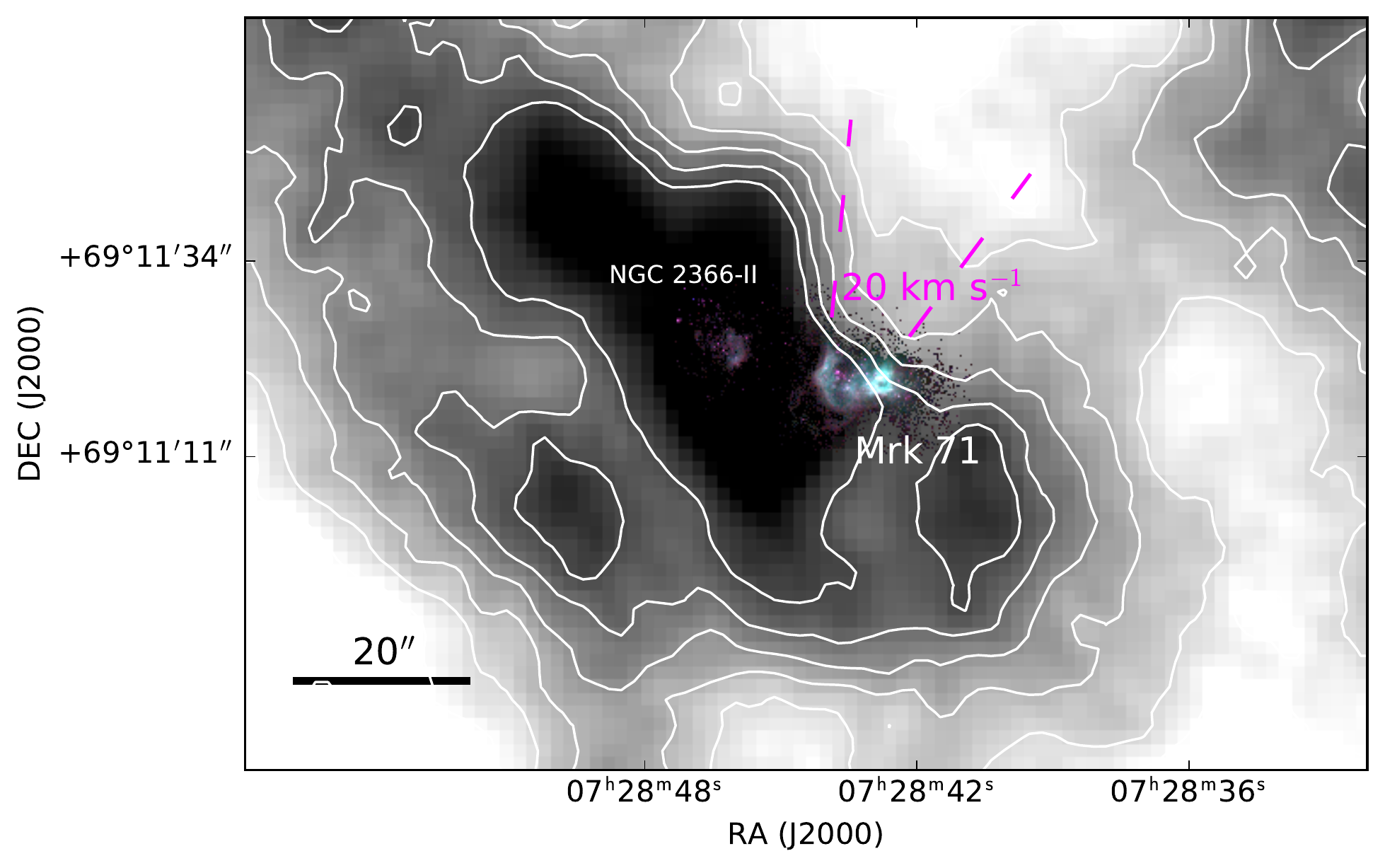}
\caption{\mk~three-color image superimposed on a gray scale, integrated $N$(\hi) map of the NGC 2366
  system~\citep{2008AJ....136.2563W}. The \hi\ map has major and minor axes of the synthesized beam of $13.1''$ and $11.85''$, respectively, a channel width of $\Delta v=2.6$ km/s, and a total bandwidth of $1.56$ MHz. Overplotted are \hi\ column density contours between $1\mbox{--}5\times10^{21}$ cm${}^{-2}$, in bins of $0.5$. The location of the $20$ km s${}^{-1}$ blowout region from \citet{1991ApJ...367..141R} is also indicated. The RGB bands are the same as in Figure~\ref{fig:rgb}. \protect\label{fig:HI}} 
\end{figure*}

Ionization-parameter mapping~\citep[IPM, ][]{2012ApJ...755...40P} 
can help diagnose density-bounded, optically thin nebulae using
spatially resolved emission-line data. In Figure~\ref{fig:rgb} we use
[O II]$\lambda3727$, [O III]$\lambda 5007$, and He II
$\lambda4686$ as the red, green, and blue channels, respectively, of
the 3-color image.  Figure~\ref{fig:rgb} shows that the central, highly
ionized region (\oiii; green) is bounded by lower-ionization zones in all directions
except to the north, where we see unbounded strong emission in
\oiii\ in the direction of the blowout described above (Figure~\ref{fig:HI}).
In all other directions, a transition zone to 
lower ionization (\oii; red) and neutral gas is visible in Figure~\ref{fig:rgb}.
This implies that the region is probably
optically thin in the \lyc, along the blowout.
The nebular morphology also strongly suggests that
the optically thin region corresponds to an ionized outflow.
The ionization structure and morphology is consistent with the
extreme nebular excitation and ionization parameter implied by the
\oiii/\oii\ ratio and presence of \heii\ $\lambda4686$ (\S~\ref{sec:excitation}). As discussed in \S~\ref{sec:burstpop}, VMS stars may be responsible
for these properties. 

The detection of C III] $\lambda1909$ emission in \mk\ may also indicate conditions favorable to \lyc~escape. Both observations~\citep[][]{2014MNRAS.445.3200S} and models~\citep{2016ApJ...833..136J} show that galaxies with strong C III] have high ionization parameters, low metallicities and young stellar populations. \citet{2016ApJ...833..136J} suggest that C III] $\lambda1909$ emission that is lower than predicted for a given age and metallicity might indicate \lyc~escape. These authors estimate that a decrease in the strength of C III] of $2\mbox{--}68\%$ could imply up to $20\%$ escape of \lyc~photons. If taken at face value, the equivalent width we estimate from the available data, EW(C III])$=14.5\pm2.0$ \AA, is lower than the predicted model value by $\sim12\%$, which would be consistent with \lyc~escape.  Within the uncertainty, however, the data are also marginally consistent with the expected value from a $\log U=-2$ model. We note that another strong local C III] emitter is Tol 1214--277, which has an equivalent width similar to \mk\ \citep{2015ApJ...814L...6R}, and is suspected of leaking \lyc\ based on analysis of its peculiar Ly$\alpha$ profile~\citep{2015A&A...578A...7V}.  

While Figure~\ref{fig:HI} shows detection of H I toward \mk, it is
likely that clumping of this gas in the foreground may also allow for
an optically thin line of sight.  In a ``picket fence''
model~\citep{2001ApJ...558...56H,Bergvall2006}, an optically thick
\hi\ medium with covering fraction less than unity allows \lyc~photons
to escape unimpeded through low density channels. It  also seems likely
that most of the detected H I is located behind \mk. As described in
\S \ref{sec:coldens}, low absorption columns for Knot A are suggested
by the optically thin Si II $\lambda1260$/$\lambda1526$ ratio and
weak detections or non-detections of other low-ionization and neutral
species. The low reddening (\S~\ref{sec:dust}) is further consistent
with these conditions, facilitating the escape of \lyc\ emission.  

Thus, we conclude that \ngc, and in particular \mk, is an excellent candidate for \lyc~emission
and could provide important clues to the conditions and processes
for the escape of ionizing radiation in GPs and higher-redshift
galaxies.  The wealth of spatially resolved data for \mk\ show 
a major, two-stage starburst, probably triggered by the interaction
between NGC~2366 and NGC~2363.  The Knot B SSC was
formed 3 -- 5 Myr ago, generating a superbubble and blowout from the
plane of the host galaxy.  The Knot A SSC
formed nearby, $\lesssim 1$ Myr ago, and while still enshrouded, its
powerful luminosity drives the high ionization parameter and other extreme
excitation properties of the \mk\ complex. The synergy between the
feedback from the older Knot B and the younger
Knot A may be one recipe for enabling \lyc~escape from this galaxy; 
similar two-stage starbursts are seen in other SSC systems, for example,
the dual system of R136 and Hodge 301 in 30 Dor \citep[e.g.,][]{2016ApJS..222...11S}.
The similarity to the integrated properties of GPs 
suggest that such processes may also be important in those objects and
other LCE candidates.  We will report a more detailed analysis of
the physical processes in \mk\ in a forthcoming work.

\subsection{A faint local population of LCEs?}

We noted earlier that an important difference between \ngc\ and the GPs is that the former is much fainter than average, and its young stellar population is much less massive (\S~\ref{sec:bigGP}). Coincidentally, the cosmic star-formation rate density decreases by a factor of $\sim0.5$ between $z=0.2\mbox{--}0.4$ and the present~\citep[e.g.][]{2014ARA&A..52..415M}, which is about the same as the difference in sSFR between the GPs and \mk/\ngc.

The system is $1\mbox{--}2$ orders of magnitude fainter in FUV than typically observed for the average GPs sample. Therefore, \ngc~probes an unexplored region in the GP mass and luminosity function. The detection of a GP analog and LCE candidate at a distance of only $3.4$ Mpc argues against the rarity of faint GP-like objects, and thus faint LCE candidates may be common. Objects with similar properties to \ngc, and in particular \mk, could be numerous at higher redshifts but remain undetected. Indeed, it would take high magnification for such objects to be detected at high redshift, as recently done by~\citet{Vanzella2017}, who identify two lensed low-mass ($<10^7$M${}_{\sun}$), low-metallicity ($1/10$ solar) compact objects at $z=3.2$, showing optical oxygen line ratios consistent with extreme GPs. If such objects indeed translate into a substantial LCE population, it would be consistent with the growing evidence that extremely faint galaxies contribute critically to the reionization of the Universe~\citep[e.g.][]{2010ApJ...710.1239R,2012ApJ...759L..38A,2015ApJ...811..140B,2015ApJ...814...69A,2016ARA&A..54..761S,2016ApJ...828...71D}. In general, these works suggest that low-luminosity LCEs should have higher LyC escape fractions. Our crude estimate of the blowout opening angle in \S~\ref{sec:clusterB} suggests, at face value, a LyC escape fraction from \ngc\ of $\lesssim 2$\%, which is similar to values seen in other confirmed local LCEs.  However, as discussed above, the LIS absorption lines hint that the optical depth in the line of sight may also be low, suggesting the possible existence of multiple channels for the escape of ionizing photons.  We suggest that \ngc\ may be representative of such a population of faint LCE galaxies at high redshift.  As such, it offers a unique opportunity to investigate in detail the properties of such objects and mechanisms for \lyc\ escape. 

\section{Conclusion}\protect\label{sec:conclude}
We have presented a comprehensive, quantitative comparison of the properties of \ngc\ and its giant H II region, \mk, with Green Pea galaxies, and find that they are in remarkable agreement.  As summarized in Table~\ref{tab:comparison}, \ngc, dominated by \mk, is quantitatively similar to GPs in almost all its properties, including very high excitation, strong ${\rm EW}(\rm [OIII])\sim2200\pm300$ \AA, extreme line ratios of $[\rm OIII]\lambda\lambda 5007,4959/[\rm O II]\lambda3727\sim17$, nebular $T_e\sim 15,000$K and $n_e\sim 200$ cm${}^{-3}$. Like in the GPs, the resulting high ionization parameter of $\log U\sim -2.2$ is generated primarily by an extremely young burst, perhaps $\lesssim 1$ Myr old, corresponding to a high sSFR $\sim5\times10^{-10}$ yr${}^{-1}$ and compact morphology.  Other similar properties include high-velocity gas kinematics (FWHM$\sim2400$ km s${}^{-1}$) and Doppler shifts indicative of outflow, low reddening ($C($\hb$)\sim0.13$), and low implied neutral gas absorption columns. Thus \ngc\ is an outstanding GP analog.  

The GP class is known for offering strong LCE candidates, some of which are spectroscopically confirmed. Indeed, we also find compelling evidence of possible \lyc~escape from our newly identified GP analog, \mk/\ngc. The wealth of existing data for this system, including HST nebular imaging \citep[e.g.,][]{2016ApJ...816...40J}, spatially resolved, nebular kinematic data \citep[e.g.,][]{1991ApJ...367..141R}, and \hi\ mapping \citep[e.g.,][]{2012AJ....144..134H} show a clear blowout and outflow to the north, generated by the older of \mk's two super star clusters, Knot B. Ionization-parameter mapping indicates that this blowout region is optically thin, with the ionization dominated by the extremely young, still enshrouded super star cluster, Knot A, whose ionizing luminosity is an order of magnitude greater than that of Knot B. Knot A is a remarkable object, with strong similarities to R136 in 30 Dor and NGC 5253 \#5, and it may contain VMS stars. This object may also be optically thin to the LyC in our line of sight, as suggested by weak detection or non-detection of low-ionization species in absorption, for example, C II $\lambda1335$, Fe II $\lambda1608, \lambda2370, \lambda2600$, Mg II $2800$~\citep{2011AJ....141...37L}, and Na I $\lambda\lambda 5890,5896$ \citep{2004ApJ...610..201S}. Si II is optically thin, as indicated by the ratio of Si II$ \lambda1260$ / Si II $\lambda1526$ \citep{2011AJ....141...37L}. The C III]$\lambda1909$ emission also may be lower than expected, and suggestive of optically thin conditions \citep{2016ApJ...833..136J}. These species all suggest low neutral column densities, consistent with the likelihood of \lyc~escape, and consistent with similar observations of GP LCE candidates.  Also similar to GPs, observed low dust reddening in \mk\ and extreme excitation conditions further favor the escape of LyC radiation.  

\mk\ is however two orders of magnitude less luminous than average and extreme GPs. The mass ($2.6\times10^{8}$ M${}_{\sun}$) and metallicity ($12+{\rm log}({\rm O}/{\rm H})=7.9$) of its host galaxy, \ngc, are at the low end of the reported GP mass and metallicity ranges~\citep{Izotov2011}. Thus, if this system is indeed an LCE, its proximity further suggests that faint LCEs may be commonplace, supporting a significant role in cosmic reionization. \mk/\ngc, as a Green Pea analog at a distance of only $3.4$ Mpc and LCE candidate, offers an unprecedentedly detailed look at the morphology and physical conditions of a potential \lyc~emitter.     

\acknowledgments
We thank Nils Bergvall, Norberto Castro, Charles Cowley, Jim Dale,
Sergiy Silich, and Linda Smith for helpful discussions.  We also thank
Kim Sokal for providing us with spectra of Knot A and Knot B.  We are
also grateful to the anonymous referee for helpful suggestions.



\vspace{5mm}





\bibliographystyle{aasjournal.bst}
\bibliography{ngc2366}

\begin{thebibliography}{}
\expandafter\ifx\csname natexlab\endcsname\relax\def\natexlab#1{#1}\fi

\bibitem[{{Alvarez} {et~al.}(2012){Alvarez}, {Finlator}, \&
  {Trenti}}]{2012ApJ...759L..38A}
{Alvarez}, M.~A., {Finlator}, K., \& {Trenti}, M. 2012, \apjl, 759, L38

\bibitem[{{Amor{\'{\i}}n} {et~al.}(2012{\natexlab{a}}){Amor{\'{\i}}n},
  {P{\'e}rez-Montero}, {V{\'{\i}}lchez}, \& {Papaderos}}]{2012ApJ...749..185A}
{Amor{\'{\i}}n}, R., {P{\'e}rez-Montero}, E., {V{\'{\i}}lchez}, J.~M., \&
  {Papaderos}, P. 2012{\natexlab{a}}, \apj, 749, 185

\bibitem[{{Amor{\'{\i}}n} {et~al.}(2012{\natexlab{b}}){Amor{\'{\i}}n},
  {V{\'{\i}}lchez}, {H{\"a}gele}, {Firpo}, {P{\'e}rez-Montero}, \&
  {Papaderos}}]{2012ApJ...754L..22A}
{Amor{\'{\i}}n}, R., {V{\'{\i}}lchez}, J.~M., {H{\"a}gele}, G.~F., {et~al.}
  2012{\natexlab{b}}, \apjl, 754, L22

\bibitem[{{Amor{\'{\i}}n} {et~al.}(2010){Amor{\'{\i}}n}, {P{\'e}rez-Montero},
  \& {V{\'{\i}}lchez}}]{2010ApJ...715L.128A}
{Amor{\'{\i}}n}, R.~O., {P{\'e}rez-Montero}, E., \& {V{\'{\i}}lchez}, J.~M.
  2010, \apjl, 715, L128

\bibitem[{{Atek} {et~al.}(2015){Atek}, {Richard}, {Jauzac}, {Kneib},
  {Natarajan}, {Limousin}, {Schaerer}, {Jullo}, {Ebeling}, {Egami}, \&
  {Clement}}]{2015ApJ...814...69A}
{Atek}, H., {Richard}, J., {Jauzac}, M., {et~al.} 2015, \apj, 814, 69

\bibitem[{{Baldwin} {et~al.}(1981){Baldwin}, {Phillips}, \&
  {Terlevich}}]{1981PASP...93....5B}
{Baldwin}, J.~A., {Phillips}, M.~M., \& {Terlevich}, R. 1981, \pasp, 93, 5

\bibitem[{{Bergvall} {et~al.}(2006){Bergvall}, {Zackrisson}, {Andersson},
  {Arnberg}, {Masegosa}, \& {{\"O}stlin}}]{Bergvall2006}
{Bergvall}, N., {Zackrisson}, E., {Andersson}, B.-G., {et~al.} 2006, \aap, 448,
  513

\bibitem[{{Binette} {et~al.}(2009){Binette}, {Drissen}, {Ubeda}, {Raga},
  {Robert}, \& {Krongold}}]{2009AA...500..817B}
{Binette}, L., {Drissen}, L., {Ubeda}, L., {et~al.} 2009, \aap, 500, 817

\bibitem[{{Borthakur} {et~al.}(2014){Borthakur}, {Heckman}, {Leitherer}, \&
  {Overzier}}]{2014Sci...346..216B}
{Borthakur}, S., {Heckman}, T.~M., {Leitherer}, C., \& {Overzier}, R.~A. 2014,
  Science, 346, 216

\bibitem[{{Bouwens} {et~al.}(2015){Bouwens}, {Illingworth}, {Oesch}, {Caruana},
  {Holwerda}, {Smit}, \& {Wilkins}}]{2015ApJ...811..140B}
{Bouwens}, R.~J., {Illingworth}, G.~D., {Oesch}, P.~A., {et~al.} 2015, \apj,
  811, 140

\bibitem[{{Calzetti} {et~al.}(2000){Calzetti}, {Armus}, {Bohlin}, {Kinney},
  {Koornneef}, \& {Storchi-Bergmann}}]{2000ApJ...533..682C}
{Calzetti}, D., {Armus}, L., {Bohlin}, R.~C., {et~al.} 2000, \apj, 533, 682

\bibitem[{{Cardamone} {et~al.}(2009){Cardamone}, {Schawinski}, {Sarzi},
  {Bamford}, {Bennert}, {Urry}, {Lintott}, {Keel}, {Parejko}, {Nichol},
  {Thomas}, {Andreescu}, {Murray}, {Raddick}, {Slosar}, {Szalay}, \&
  {Vandenberg}}]{2009MNRAS.399.1191C}
{Cardamone}, C., {Schawinski}, K., {Sarzi}, M., {et~al.} 2009, \mnras, 399,
  1191

\bibitem[{{Corwin}(2006)}]{Corwin2006}
{Corwin}, H.~G. 2006, {Historical Notes: NGC 2000 thru NGC 2999 (Aug, 2006)},
  {The NGC/IC Project},
  \url{http://www.ngcicproject.org/corwin/DataFiles/Aug_2006/ngcnotes_2.txt},
  Accessed: 2016, March 1

\bibitem[{{Cox} {et~al.}(2007){Cox}, {Cordiner}, {Ehrenfreund}, {Kaper},
  {Sarre}, {Foing}, {Spaans}, {Cami}, {Sofia}, {Clayton}, {Gordon}, \&
  {Salama}}]{2007A&A...470..941C}
{Cox}, N.~L.~J., {Cordiner}, M.~A., {Ehrenfreund}, P., {et~al.} 2007, \aap,
  470, 941

\bibitem[{{Crowther} {et~al.}(2010){Crowther}, {Schnurr}, {Hirschi}, {Yusof},
  {Parker}, {Goodwin}, \& {Kassim}}]{2010MNRAS.408..731C}
{Crowther}, P.~A., {Schnurr}, O., {Hirschi}, R., {et~al.} 2010, \mnras, 408,
  731

\bibitem[{{Crowther} \& {Walborn}(2011)}]{2011MNRAS.416.1311C}
{Crowther}, P.~A., \& {Walborn}, N.~R. 2011, \mnras, 416, 1311

\bibitem[{{Dale} {et~al.}(2009){Dale}, {Cohen}, {Johnson}, {Schuster},
  {Calzetti}, {Engelbracht}, {Gil de Paz}, {Kennicutt}, {Lee}, {Begum},
  {Block}, {Dalcanton}, {Funes}, {Gordon}, {Johnson}, {Marble}, {Sakai},
  {Skillman}, {van Zee}, {Walter}, {Weisz}, {Williams}, {Wu}, \&
  {Wu}}]{2009ApJ...703..517D}
{Dale}, D.~A., {Cohen}, S.~A., {Johnson}, L.~C., {et~al.} 2009, \apj, 703, 517

\bibitem[{{de Barros} {et~al.}(2016){de Barros}, {Vanzella}, {Amor{\'{\i}}n},
  {Castellano}, {Siana}, {Grazian}, {Suh}, {Balestra}, {Vignali}, {Verhamme},
  {Zamorani}, {Mignoli}, {Hasinger}, {Comastri}, {Pentericci},
  {P{\'e}rez-Montero}, {Fontana}, {Giavalisco}, \&
  {Gilli}}]{2016A&A...585A..51D}
{de Barros}, S., {Vanzella}, E., {Amor{\'{\i}}n}, R., {et~al.} 2016, \aap, 585,
  A51

\bibitem[{{de Vaucouleurs} {et~al.}(1991){de Vaucouleurs}, {de Vaucouleurs},
  {Corwin}, {Buta}, {Paturel}, \& {Fouqu{\'e}}}]{1991rc3..book.....D}
{de Vaucouleurs}, G., {de Vaucouleurs}, A., {Corwin}, Jr., H.~G., {et~al.}
  1991, {Third Reference Catalogue of Bright Galaxies.} (New York, NY, USA:
  Springer)

\bibitem[{{Dijkstra} {et~al.}(2016){Dijkstra}, {Gronke}, \&
  {Venkatesan}}]{2016ApJ...828...71D}
{Dijkstra}, M., {Gronke}, M., \& {Venkatesan}, A. 2016, \apj, 828, 71

\bibitem[{{Drissen} {et~al.}(1993){Drissen}, {Roy}, \&
  {Moffat}}]{1993AJ....106.1460D}
{Drissen}, L., {Roy}, J.-R., \& {Moffat}, A.~F.~J. 1993, \aj, 106, 1460

\bibitem[{{Drissen} {et~al.}(2000){Drissen}, {Roy}, {Robert}, {Devost}, \&
  {Doyon}}]{2000AJ....119..688D}
{Drissen}, L., {Roy}, J.-R., {Robert}, C., {Devost}, D., \& {Doyon}, R. 2000,
  \aj, 119, 688

\bibitem[{{Dufour} \& {Harlow}(1977)}]{1977ApJ...216..706D}
{Dufour}, R.~J., \& {Harlow}, W.~V. 1977, \apj, 216, 706

\bibitem[{{Esteban} {et~al.}(2002){Esteban}, {Peimbert}, {Torres-Peimbert}, \&
  {Rodr{\'{\i}}guez}}]{2002ApJ...581..241E}
{Esteban}, C., {Peimbert}, M., {Torres-Peimbert}, S., \& {Rodr{\'{\i}}guez}, M.
  2002, \apj, 581, 241

\bibitem[{{Ferland} {et~al.}(1998){Ferland}, {Korista}, {Verner}, {Ferguson},
  {Kingdon}, \& {Verner}}]{1998PASP..110..761F}
{Ferland}, G.~J., {Korista}, K.~T., {Verner}, D.~A., {et~al.} 1998, \pasp, 110,
  761

\bibitem[{{Fernandez} \& {Shull}(2011)}]{2011ApJ...731...20F}
{Fernandez}, E.~R., \& {Shull}, J.~M. 2011, \apj, 731, 20

\bibitem[{{Garnett} {et~al.}(1995){Garnett}, {Dufour}, {Peimbert},
  {Torres-Peimbert}, {Shields}, {Skillman}, {Terlevich}, \&
  {Terlevich}}]{Garnett1995}
{Garnett}, D.~R., {Dufour}, R.~J., {Peimbert}, M., {et~al.} 1995, \apjl, 449,
  L77

\bibitem[{{Gonz{\'a}lez Delgado} {et~al.}(1999){Gonz{\'a}lez Delgado},
  {Leitherer}, \& {Heckman}}]{1999ApJS..125..489G}
{Gonz{\'a}lez Delgado}, R.~M., {Leitherer}, C., \& {Heckman}, T.~M. 1999,
  \apjs, 125, 489

\bibitem[{{Gonzalez-Delgado} {et~al.}(1994){Gonzalez-Delgado}, {Perez},
  {Tenorio-Tagle}, {Vilchez}, {Terlevich}, {Terlevich}, {Telles},
  {Rodriguez-Espinosa}, {Mas-Hesse}, {Garcia-Vargas}, {Diaz}, {Cepa}, \&
  {Castaneda}}]{1994ApJ...437..239G}
{Gonzalez-Delgado}, R.~M., {Perez}, E., {Tenorio-Tagle}, G., {et~al.} 1994,
  \apj, 437, 239

\bibitem[{{Gr{\"a}fener} \& {Vink}(2015)}]{2015A&A...578L...2G}
{Gr{\"a}fener}, G., \& {Vink}, J.~S. 2015, \aap, 578, L2

\bibitem[{{Guaita} {et~al.}(2016){Guaita}, {Pentericci}, {Grazian}, {Vanzella},
  {Nonino}, {Giavalisco}, {Zamorani}, {Bongiorno}, {Cassata}, {Castellano},
  {Garilli}, {Gawiser}, {Le Brun}, {Le F{\`e}vre}, {Lemaux}, {Maccagni},
  {Merlin}, {Santini}, {Tasca}, {Thomas}, {Zucca}, {De Barros}, {Hathi},
  {Amorin}, {Bardelli}, \& {Fontana}}]{2016A&A...587A.133G}
{Guaita}, L., {Pentericci}, L., {Grazian}, A., {et~al.} 2016, \aap, 587, A133

\bibitem[{{Hawley}(2012)}]{2012PASP..124...21H}
{Hawley}, S.~A. 2012, \pasp, 124, 21

\bibitem[{{Heckman} {et~al.}(2001){Heckman}, {Sembach}, {Meurer}, {Leitherer},
  {Calzetti}, \& {Martin}}]{2001ApJ...558...56H}
{Heckman}, T.~M., {Sembach}, K.~R., {Meurer}, G.~R., {et~al.} 2001, \apj, 558,
  56

\bibitem[{{Heckman} {et~al.}(2005){Heckman}, {Hoopes}, {Seibert}, {Martin},
  {Salim}, {Rich}, {Kauffmann}, {Charlot}, {Barlow}, {Bianchi}, {Byun},
  {Donas}, {Forster}, {Friedman}, {Jelinsky}, {Lee}, {Madore}, {Malina},
  {Milliard}, {Morrissey}, {Neff}, {Schiminovich}, {Siegmund}, {Small},
  {Szalay}, {Welsh}, \& {Wyder}}]{2005ApJ...619L..35H}
{Heckman}, T.~M., {Hoopes}, C.~G., {Seibert}, M., {et~al.} 2005, \apjl, 619,
  L35

\bibitem[{{Heckman} {et~al.}(2011){Heckman}, {Borthakur}, {Overzier},
  {Kauffmann}, {Basu-Zych}, {Leitherer}, {Sembach}, {Martin}, {Rich},
  {Schiminovich}, \& {Seibert}}]{2011ApJ...730....5H}
{Heckman}, T.~M., {Borthakur}, S., {Overzier}, R., {et~al.} 2011, \apj, 730, 5

\bibitem[{{Henry} {et~al.}(2015){Henry}, {Scarlata}, {Martin}, \&
  {Erb}}]{2015ApJ...809...19H}
{Henry}, A., {Scarlata}, C., {Martin}, C.~L., \& {Erb}, D. 2015, \apj, 809, 19

\bibitem[{{Hoopes} {et~al.}(2007){Hoopes}, {Heckman}, {Salim}, {Seibert},
  {Tremonti}, {Schiminovich}, {Rich}, {Martin}, {Charlot}, {Kauffmann},
  {Forster}, {Friedman}, {Morrissey}, {Neff}, {Small}, {Wyder}, {Bianchi},
  {Donas}, {Lee}, {Madore}, {Milliard}, {Szalay}, {Welsh}, \&
  {Yi}}]{2007ApJS..173..441H}
{Hoopes}, C.~G., {Heckman}, T.~M., {Salim}, S., {et~al.} 2007, \apjs, 173, 441

\bibitem[{{Hunter} {et~al.}(2001){Hunter}, {Elmegreen}, \& {van
  Woerden}}]{2001ApJ...556..773H}
{Hunter}, D.~A., {Elmegreen}, B.~G., \& {van Woerden}, H. 2001, \apj, 556, 773

\bibitem[{{Hunter} \& {Hoffman}(1999)}]{1999AJ....117.2789H}
{Hunter}, D.~A., \& {Hoffman}, L. 1999, \aj, 117, 2789

\bibitem[{{Hunter} {et~al.}(2012){Hunter}, {Ficut-Vicas}, {Ashley}, {Brinks},
  {Cigan}, {Elmegreen}, {Heesen}, {Herrmann}, {Johnson}, {Oh}, {Rupen},
  {Schruba}, {Simpson}, {Walter}, {Westpfahl}, {Young}, \&
  {Zhang}}]{2012AJ....144..134H}
{Hunter}, D.~A., {Ficut-Vicas}, D., {Ashley}, T., {et~al.} 2012, \aj, 144, 134

\bibitem[{{Izotov} {et~al.}(2011){Izotov}, {Guseva}, {Fricke}, \&
  {Henkel}}]{2011A&A...536L...7I}
{Izotov}, Y.~I., {Guseva}, N.~G., {Fricke}, K.~J., \& {Henkel}, C. 2011, \aap,
  536, L7

\bibitem[{{Izotov} {et~al.}(2014){Izotov}, {Guseva}, {Fricke}, \&
  {Henkel}}]{2014A&A...561A..33I}
---. 2014, \aap, 561, A33

\bibitem[{Izotov {et~al.}(2011)Izotov, Guseva, \& Thuan}]{Izotov2011}
Izotov, Y.~I., Guseva, N.~G., \& Thuan, T.~X. 2011, \apj, 728, 161

\bibitem[{{Izotov} {et~al.}(2016{\natexlab{a}}){Izotov}, {Orlitov{\'a}},
  {Schaerer}, {Thuan}, {Verhamme}, {Guseva}, \&
  {Worseck}}]{2016Natur.529..178I}
{Izotov}, Y.~I., {Orlitov{\'a}}, I., {Schaerer}, D., {et~al.}
  2016{\natexlab{a}}, \nat, 529, 178

\bibitem[{{Izotov} {et~al.}(2016{\natexlab{b}}){Izotov}, {Schaerer}, {Thuan},
  {Worseck}, {Guseva}, {Orlitov{\'a}}, \& {Verhamme}}]{2016MNRAS.461.3683I}
{Izotov}, Y.~I., {Schaerer}, D., {Thuan}, T.~X., {et~al.} 2016{\natexlab{b}},
  \mnras, 461, 3683

\bibitem[{{Izotov} \& {Thuan}(2011)}]{2011ApJ...734...82I}
{Izotov}, Y.~I., \& {Thuan}, T.~X. 2011, \apj, 734, 82

\bibitem[{{Izotov} {et~al.}(1997){Izotov}, {Thuan}, \&
  {Lipovetsky}}]{1997ApJS..108....1I}
{Izotov}, Y.~I., {Thuan}, T.~X., \& {Lipovetsky}, V.~A. 1997, \apjs, 108, 1

\bibitem[{{James} {et~al.}(2014){James}, {Aloisi}, {Heckman}, {Sohn}, \&
  {Wolfe}}]{2014ApJ...795..109J}
{James}, B.~L., {Aloisi}, A., {Heckman}, T., {Sohn}, S.~T., \& {Wolfe}, M.~A.
  2014, \apj, 795, 109

\bibitem[{{James} {et~al.}(2016){James}, {Auger}, {Aloisi}, {Calzetti}, \&
  {Kewley}}]{2016ApJ...816...40J}
{James}, B.~L., {Auger}, M., {Aloisi}, A., {Calzetti}, D., \& {Kewley}, L.
  2016, \apj, 816, 40

\bibitem[{{James} {et~al.}(2013){James}, {Tsamis}, {Walsh}, {Barlow}, \&
  {Westmoquette}}]{2013MNRAS.430.2097J}
{James}, B.~L., {Tsamis}, Y.~G., {Walsh}, J.~R., {Barlow}, M.~J., \&
  {Westmoquette}, M.~S. 2013, \mnras, 430, 2097

\bibitem[{{Jaskot} \& {Oey}(2013)}]{2013ApJ...766...91J}
{Jaskot}, A.~E., \& {Oey}, M.~S. 2013, \apj, 766, 91

\bibitem[{{Jaskot} \& {Oey}(2014)}]{2014ApJ...791L..19J}
---. 2014, \apjl, 791, L19

\bibitem[{{Jaskot} \& {Ravindranath}(2016)}]{2016ApJ...833..136J}
{Jaskot}, A.~E., \& {Ravindranath}, S. 2016, \apj, 833, 136

\bibitem[{{Kennicutt} {et~al.}(1980){Kennicutt}, {Balick}, \&
  {Heckman}}]{1980PASP...92..134K}
{Kennicutt}, R., {Balick}, B., \& {Heckman}, T. 1980, \pasp, 92, 134

\bibitem[{{Kennicutt} {et~al.}(2008){Kennicutt}, {Lee}, {Funes}, {J.}, {Sakai},
  \& {Akiyama}}]{2008ApJS..178..247K}
{Kennicutt}, Jr., R.~C., {Lee}, J.~C., {Funes}, J.~G., {et~al.} 2008, \apjs,
  178, 247

\bibitem[{{Leitet} {et~al.}(2013){Leitet}, {Bergvall}, {Hayes}, {Linn{\'e}}, \&
  {Zackrisson}}]{2013A&A...553A.106L}
{Leitet}, E., {Bergvall}, N., {Hayes}, M., {Linn{\'e}}, S., \& {Zackrisson}, E.
  2013, \aap, 553, A106

\bibitem[{{Leitet} {et~al.}(2011){Leitet}, {Bergvall}, {Piskunov}, \&
  {Andersson}}]{2011A&A...532A.107L}
{Leitet}, E., {Bergvall}, N., {Piskunov}, N., \& {Andersson}, B.-G. 2011, \aap,
  532, A107

\bibitem[{{Leitherer} {et~al.}(2014){Leitherer}, {Ekstr{\"o}m}, {Meynet},
  {Schaerer}, {Agienko}, \& {Levesque}}]{2014ApJS..212...14L}
{Leitherer}, C., {Ekstr{\"o}m}, S., {Meynet}, G., {et~al.} 2014, \apjs, 212, 14

\bibitem[{{Leitherer} {et~al.}(1995){Leitherer}, {Ferguson}, {Heckman}, \&
  {Lowenthal}}]{1995ApJ...454L..19L}
{Leitherer}, C., {Ferguson}, H.~C., {Heckman}, T.~M., \& {Lowenthal}, J.~D.
  1995, \apjl, 454, L19

\bibitem[{{Leitherer} {et~al.}(2016){Leitherer}, {Hernandez}, {Lee}, \&
  {Oey}}]{2016ApJ...823...64L}
{Leitherer}, C., {Hernandez}, S., {Lee}, J.~C., \& {Oey}, M.~S. 2016, \apj,
  823, 64

\bibitem[{{Leitherer} {et~al.}(2011){Leitherer}, {Tremonti}, {Heckman}, \&
  {Calzetti}}]{2011AJ....141...37L}
{Leitherer}, C., {Tremonti}, C.~A., {Heckman}, T.~M., \& {Calzetti}, D. 2011,
  \aj, 141, 37

\bibitem[{{Lelli} {et~al.}(2014){Lelli}, {Verheijen}, \&
  {Fraternali}}]{2014AA...566A..71L}
{Lelli}, F., {Verheijen}, M., \& {Fraternali}, F. 2014, \aap, 566, A71

\bibitem[{{Luridiana} {et~al.}(1999){Luridiana}, {Peimbert}, \&
  {Leitherer}}]{1999ApJ...527..110L}
{Luridiana}, V., {Peimbert}, M., \& {Leitherer}, C. 1999, \apj, 527, 110

\bibitem[{{Madau} \& {Dickinson}(2014)}]{2014ARA&A..52..415M}
{Madau}, P., \& {Dickinson}, M. 2014, \araa, 52, 415

\bibitem[{{Masegosa} {et~al.}(1991){Masegosa}, {Moles}, \& {del
  Olmo}}]{Masegosa1991}
{Masegosa}, J., {Moles}, M., \& {del Olmo}, A. 1991, \aap, 249, 505

\bibitem[{{McQuinn} {et~al.}(2015){McQuinn}, {Skillman}, {Dolphin}, \&
  {Mitchell}}]{2015ApJ...808..109M}
{McQuinn}, K.~B.~W., {Skillman}, E.~D., {Dolphin}, A.~E., \& {Mitchell}, N.~P.
  2015, \apj, 808, 109

\bibitem[{{McQuinn} {et~al.}(2010){McQuinn}, {Skillman}, {Cannon}, {Dalcanton},
  {Dolphin}, {Hidalgo-Rodr{\'{\i}}guez}, {Holtzman}, {Stark}, {Weisz}, \&
  {Williams}}]{2010ApJ...721..297M}
{McQuinn}, K.~B.~W., {Skillman}, E.~D., {Cannon}, J.~M., {et~al.} 2010, \apj,
  721, 297

\bibitem[{{Micheva} {et~al.}(2017){Micheva}, {Iwata}, {Inoue}, {Matsuda},
  {Yamada}, \& {Hayashino}}]{2017MNRAS.465..316M}
{Micheva}, G., {Iwata}, I., {Inoue}, A.~K., {et~al.} 2017, \mnras, 465, 316

\bibitem[{{Moustakas} \& {Kennicutt}(2006)}]{2006ApJS..164...81M}
{Moustakas}, J., \& {Kennicutt}, Jr., R.~C. 2006, \apjs, 164, 81

\bibitem[{{Nakajima} \& {Ouchi}(2014)}]{2014MNRAS.442..900N}
{Nakajima}, K., \& {Ouchi}, M. 2014, \mnras, 442, 900

\bibitem[{{Noeske} {et~al.}(2000){Noeske}, {Guseva}, {Fricke}, {Izotov},
  {Papaderos}, \& {Thuan}}]{2000A&A...361...33N}
{Noeske}, K.~G., {Guseva}, N.~G., {Fricke}, K.~J., {et~al.} 2000, \aap, 361, 33

\bibitem[{{Overzier} {et~al.}(2010){Overzier}, {Heckman}, {Schiminovich},
  {Basu-Zych}, {Gon{\c c}alves}, {Martin}, \& {Rich}}]{2010ApJ...710..979O}
{Overzier}, R.~A., {Heckman}, T.~M., {Schiminovich}, D., {et~al.} 2010, \apj,
  710, 979

\bibitem[{{Overzier} {et~al.}(2008){Overzier}, {Heckman}, {Kauffmann},
  {Seibert}, {Rich}, {Basu-Zych}, {Lotz}, {Aloisi}, {Charlot}, {Hoopes},
  {Martin}, {Schiminovich}, \& {Madore}}]{2008ApJ...677...37O}
{Overzier}, R.~A., {Heckman}, T.~M., {Kauffmann}, G., {et~al.} 2008, \apj, 677,
  37

\bibitem[{{Overzier} {et~al.}(2009){Overzier}, {Heckman}, {Tremonti}, {Armus},
  {Basu-Zych}, {Gon{\c c}alves}, {Rich}, {Martin}, {Ptak}, {Schiminovich},
  {Ford}, {Madore}, \& {Seibert}}]{2009ApJ...706..203O}
{Overzier}, R.~A., {Heckman}, T.~M., {Tremonti}, C., {et~al.} 2009, \apj, 706,
  203

\bibitem[{{Peimbert}(2003)}]{2003ApJ...584..735P}
{Peimbert}, A. 2003, \apj, 584, 735

\bibitem[{{Pellegrini} {et~al.}(2012){Pellegrini}, {Oey}, {Winkler}, {Points},
  {Smith}, {Jaskot}, \& {Zastrow}}]{2012ApJ...755...40P}
{Pellegrini}, E.~W., {Oey}, M.~S., {Winkler}, P.~F., {et~al.} 2012, \apj, 755,
  40

\bibitem[{{Pettini} \& {Pagel}(2004)}]{2004MNRAS.348L..59P}
{Pettini}, M., \& {Pagel}, B.~E.~J. 2004, \mnras, 348, L59

\bibitem[{{Razoumov} \& {Sommer-Larsen}(2010)}]{2010ApJ...710.1239R}
{Razoumov}, A.~O., \& {Sommer-Larsen}, J. 2010, \apj, 710, 1239

\bibitem[{{Rigby} {et~al.}(2015){Rigby}, {Bayliss}, {Gladders}, {Sharon},
  {Wuyts}, {Dahle}, {Johnson}, \& {Pe{\~n}a-Guerrero}}]{2015ApJ...814L...6R}
{Rigby}, J.~R., {Bayliss}, M.~B., {Gladders}, M.~D., {et~al.} 2015, \apjl, 814,
  L6

\bibitem[{{Rosa} {et~al.}(1984){Rosa}, {Joubert}, \&
  {Benvenuti}}]{1984A&AS...57..361R}
{Rosa}, M., {Joubert}, M., \& {Benvenuti}, P. 1984, \aaps, 57, 361

\bibitem[{{Roy} {et~al.}(1992){Roy}, {Aube}, {McCall}, \&
  {Dufour}}]{1992ApJ...386..498R}
{Roy}, J.-R., {Aube}, M., {McCall}, M.~L., \& {Dufour}, R.~J. 1992, \apj, 386,
  498

\bibitem[{{Roy} {et~al.}(1991){Roy}, {Boulesteix}, {Joncas}, \&
  {Grundseth}}]{1991ApJ...367..141R}
{Roy}, J.-R., {Boulesteix}, J., {Joncas}, G., \& {Grundseth}, B. 1991, \apj,
  367, 141

\bibitem[{{Sabbi} {et~al.}(2016){Sabbi}, {Lennon}, {Anderson}, {Cignoni}, {van
  der Marel}, {Zaritsky}, {De Marchi}, {Panagia}, {Gouliermis}, {Grebel},
  {Gallagher}, {Smith}, {Sana}, {Aloisi}, {Tosi}, {Evans}, {Arab}, {Boyer}, {de
  Mink}, {Gordon}, {Koekemoer}, {Larsen}, {Ryon}, \&
  {Zeidler}}]{2016ApJS..222...11S}
{Sabbi}, E., {Lennon}, D.~J., {Anderson}, J., {et~al.} 2016, \apjs, 222, 11

\bibitem[{{Schwartz} \& {Martin}(2004)}]{2004ApJ...610..201S}
{Schwartz}, C.~M., \& {Martin}, C.~L. 2004, \apj, 610, 201

\bibitem[{{Shapley} {et~al.}(2016){Shapley}, {Steidel}, {Strom},
  {Bogosavljevi{\'c}}, {Reddy}, {Siana}, {Mostardi}, \&
  {Rudie}}]{2016ApJ...826L..24S}
{Shapley}, A.~E., {Steidel}, C.~C., {Strom}, A.~L., {et~al.} 2016, \apjl, 826,
  L24

\bibitem[{{Siana} {et~al.}(2007){Siana}, {Teplitz}, {Colbert}, {Ferguson},
  {Dickinson}, {Brown}, {Conselice}, {de Mello}, {Gardner}, {Giavalisco}, \&
  {Menanteau}}]{2007ApJ...668...62S}
{Siana}, B., {Teplitz}, H.~I., {Colbert}, J., {et~al.} 2007, \apj, 668, 62

\bibitem[{{Siana} {et~al.}(2015){Siana}, {Shapley}, {Kulas}, {Nestor},
  {Steidel}, {Teplitz}, {Alavi}, {Brown}, {Conselice}, {Ferguson}, {Dickinson},
  {Giavalisco}, {Colbert}, {Bridge}, {Gardner}, \& {de
  Mello}}]{2015ApJ...804...17S}
{Siana}, B., {Shapley}, A.~E., {Kulas}, K.~R., {et~al.} 2015, \apj, 804, 17

\bibitem[{{Smith} {et~al.}(2016){Smith}, {Crowther}, {Calzetti}, \&
  {Sidoli}}]{2016ApJ...823...38S}
{Smith}, L.~J., {Crowther}, P.~A., {Calzetti}, D., \& {Sidoli}, F. 2016, \apj,
  823, 38

\bibitem[{{Sokal} {et~al.}(2016){Sokal}, {Johnson}, {Indebetouw}, \&
  {Massey}}]{2016ApJ...826..194S}
{Sokal}, K.~R., {Johnson}, K.~E., {Indebetouw}, R., \& {Massey}, P. 2016, \apj,
  826, 194

\bibitem[{{Stark}(2016)}]{2016ARA&A..54..761S}
{Stark}, D.~P. 2016, \araa, 54, 761

\bibitem[{{Stark} {et~al.}(2014){Stark}, {Richard}, {Siana}, {Charlot},
  {Freeman}, {Gutkin}, {Wofford}, {Robertson}, {Amanullah}, {Watson}, \&
  {Milvang-Jensen}}]{2014MNRAS.445.3200S}
{Stark}, D.~P., {Richard}, J., {Siana}, B., {et~al.} 2014, \mnras, 445, 3200

\bibitem[{{Stasi{\'n}ska} {et~al.}(2015){Stasi{\'n}ska}, {Izotov}, {Morisset},
  \& {Guseva}}]{Stasinska2015AA}
{Stasi{\'n}ska}, G., {Izotov}, Y., {Morisset}, C., \& {Guseva}, N. 2015, \aap,
  576, A83

\bibitem[{{Steidel} {et~al.}(2001){Steidel}, {Pettini}, \&
  {Adelberger}}]{2001ApJ...546..665S}
{Steidel}, C.~C., {Pettini}, M., \& {Adelberger}, K.~L. 2001, \apj, 546, 665

\bibitem[{{Thuan} {et~al.}(2014){Thuan}, {Bauer}, \&
  {Izotov}}]{2014MNRAS.441.1841T}
{Thuan}, T.~X., {Bauer}, F.~E., \& {Izotov}, Y.~I. 2014, \mnras, 441, 1841

\bibitem[{{Thuan} \& {Izotov}(2005)}]{2005ApJ...627..739T}
{Thuan}, T.~X., \& {Izotov}, Y.~I. 2005, \apj, 627, 739

\bibitem[{{Tolstoy} {et~al.}(1995){Tolstoy}, {Saha}, {Hoessel}, \&
  {McQuade}}]{1995AJ....110.1640T}
{Tolstoy}, E., {Saha}, A., {Hoessel}, J.~G., \& {McQuade}, K. 1995, \aj, 110,
  1640

\bibitem[{{Vanzella} {et~al.}(2010){Vanzella}, {Giavalisco}, {Inoue}, {Nonino},
  {Fontanot}, {Cristiani}, {Grazian}, {Dickinson}, {Stern}, {Tozzi},
  {Giallongo}, {Ferguson}, {Spinrad}, {Boutsia}, {Fontana}, {Rosati}, \&
  {Pentericci}}]{2010ApJ...725.1011V}
{Vanzella}, E., {Giavalisco}, M., {Inoue}, A.~K., {et~al.} 2010, \apj, 725,
  1011

\bibitem[{{Vanzella} {et~al.}(2015){Vanzella}, {de Barros}, {Castellano},
  {Grazian}, {Inoue}, {Schaerer}, {Guaita}, {Zamorani}, {Giavalisco}, {Siana},
  {Pentericci}, {Giallongo}, {Fontana}, \& {Vignali}}]{2015A&A...576A.116V}
{Vanzella}, E., {de Barros}, S., {Castellano}, M., {et~al.} 2015, \aap, 576,
  A116

\bibitem[{{Vanzella} {et~al.}(2017){Vanzella}, {Castellano}, {Meneghetti},
  {Mercurio}, {Caminha}, {Cupani}, {Calura}, {Christensen}, {Merlin}, {Rosati},
  {Gronke}, {Dijkstra}, {Mignoli}, {Gilli}, {De Barros}, {Caputi}, {Grillo},
  {Balestra}, {Cristiani}, {Nonino}, {Giallongo}, {Grazian}, {Pentericci},
  {Fontana}, {Comastri}, {Vignali}, {Zamorani}, {Brusa}, {Bergamini}, \&
  {Tozzi}}]{Vanzella2017}
{Vanzella}, E., {Castellano}, M., {Meneghetti}, M., {et~al.} 2017, \apj, 842,
  47

\bibitem[{{Verhamme} {et~al.}(2015){Verhamme}, {Orlitov{\'a}}, {Schaerer}, \&
  {Hayes}}]{2015A&A...578A...7V}
{Verhamme}, A., {Orlitov{\'a}}, I., {Schaerer}, D., \& {Hayes}, M. 2015, \aap,
  578, A7

\bibitem[{{Walter} {et~al.}(2008){Walter}, {Brinks}, {de Blok}, {Bigiel},
  {Kennicutt}, {Thornley}, \& {Leroy}}]{2008AJ....136.2563W}
{Walter}, F., {Brinks}, E., {de Blok}, W.~J.~G., {et~al.} 2008, \aj, 136, 2563

\bibitem[{{Whitmore} \& {Zhang}(2002)}]{2002AJ....124.1418W}
{Whitmore}, B.~C., \& {Zhang}, Q. 2002, \aj, 124, 1418

\bibitem[{{Zastrow} {et~al.}(2013){Zastrow}, {Oey}, {Veilleux}, \&
  {McDonald}}]{2013ApJ...779...76Z}
{Zastrow}, J., {Oey}, M.~S., {Veilleux}, S., \& {McDonald}, M. 2013, \apj, 779,
  76

\end{thebibliography}

%
%



\end{document}